\documentclass[12pt]{article}
\usepackage{amsmath,amssymb,amsfonts,bm,soul}
\usepackage{fullpage,xcolor,url,graphicx,natbib}

\newtheorem{lemma}{Lemma}
\newtheorem{proposition}{Proposition}

\def\diag{{\rm diag}}
\def\pconv{\smash{\mathop{\longrightarrow}\limits^p}}     
\def\dconv{\smash{\mathop{\longrightarrow}\limits^d}}     

\newcommand{\lio}{\Lambda_i^0}
\newcommand{\fto}{F_t^0}
\newcommand{\pr}{^\prime}

\newcommand{\sumiN}{\sum_{i=1}^N}
\newcommand{\sumtT}{\sum_{t=1}^T}
\newcommand{\sumsT}{\sum_{s=1}^T}

\newcommand{\OpdNTitwo}{O_p(\delta_{NT}^{-2})}
\newcommand{\OpdNTione}{O_p(\delta_{NT}^{-1})}

\newcommand{\Fo}{\bm F^0}
\newcommand{\Fop}{\bm F^{0'}}
\newcommand{\Lo}{\bm \Lambda^0}
\newcommand{\Lop}{\bm \Lambda^{0'}}

\def\Lp{\bm \Lambda^{0\pr}}
\def\Fp{\bm F^{0\pr}}

\def\tr{\mathrm{trace}}
\newcommand{\tF}{\tilde {\bm F}}
\newcommand{\tC}{\tilde {\bm C}}
\newcommand{\tL}{\tilde {\bm \Lambda}}
\newcommand{\tFp}{\tilde {\bm F}^\prime}
\newcommand{\tLp}{\tilde {\bm \Lambda}^\prime}

\renewcommand{\a}{ {\alpha} }
\newcommand{\plim}{\mathrm{plim}}

\newcommand{\bH}{\bm H}
\newcommand{\Bi}{\bm B_N^{-1}}
\newcommand{\B}{\bm B_N}
\newcommand{\be}{\bm e}
\newcommand{\e}{\bm e}

\newcommand{\DNTr}{\bm D_{NT,r}}
\newcommand{\bX}{\bm X}
\begin{document}
\title{\large{\textsc{Approximate Factor Models with  Weaker  Loadings}}}
\author{Jushan Bai\thanks{Columbia University, 420 W. 118 St. MC 3308, New York, NY 10027.
Email: jb3064@columbia.edu}
  \and Serena Ng\thanks{Columbia University and NBER, 420 W. 118 St. MC 3308,
  New York, NY 10027. Email: serena.ng@columbia.edu.
\newline We thank an anonymous referee and an Associate Editor for helpful comments.
 This work is supported by the National Science
Foundation  SES-2018369 (Ng).}}
\date{\today\bigskip}

\maketitle
\begin{abstract}
Pervasive  cross-section dependence is increasingly recognized as a characteristic of economic data and the  approximate factor model provides a useful framework for analysis.  Assuming  a strong  factor structure where $\Lop\Lo/N^\alpha$ is positive definite in the limit when $\alpha=1$, early work established   convergence of the  principal component estimates of the  factors  and loadings up to a rotation  matrix.  This paper shows   that  the estimates are still  consistent and asymptotically normal when $\alpha\in(0,1]$  albeit at slower rates and under additional assumptions on the sample size.  The results hold whether $\alpha$ is  constant or varies across factor loadings.  The  framework developed for heterogeneous loadings  and the simplified proofs that can be also used  in  strong factor analysis are of independent interest.
\end{abstract}

JEL Classification: C30, C31

Keywords:   principal components,  low rank decomposition, weak factors, factor augmented regressions.

\thispagestyle{empty}
\setcounter{page}{0}
\newpage
\baselineskip=18.0pt

\section{Introduction}

Starting with \citet{fhlr-restat} and \citet{stock-watson-di-wp,stock-watson-jasa:02}, a large body of research has been developed to estimate the latent common variations in  large  panels in which the $N$ units observed over $T$ periods are cross-sectionally correlated.  A fundamental result shown in \citet{baing-ecta:02}  is that the space spanned by the  factors can be consistently estimated by the method of static principal components (PC) at rate $\min(\sqrt{N}, \sqrt{T})$.  \citet{bai-ecta:03} then establishes $\sqrt{N}$  asymptotic normality of the estimated factors $\tF$  up to a rotation matrix $\bm H$.   The maintained assumption is that the factor structure is strong,  meaning that if $\Fo$ and $\Lo$ are the latent factors and loadings,    the matrices $\Fop \Fo/T$ and   $\Lop \Lo/N$ are both positive definite in the limit. However,
 \citet{onatski-joe:12} shows that the PC estimates are  inconsistent when $\Lop\Lo$ (without dividing by $N$) has a positive definite limit.   This has generated a good deal of interest in  determining the number of less pervasive factors.    Some assume large idiosyncratic variances, some assume that the entries of $\Lo$  are non-zero but small, while others assume a sparse $\Lo$  with many zero entries. See, for example,  \citet{dgr-lasso},  \citet{lettau-pelger:20}, \citet{uematsu-yamagata:est}, \citet{freyaldenhoven:22}. Though the term `weak factors' is used in different ways, there is a presumption that  the PC estimator has undesirable properties  when the  strong factor assumption fails. However,  to our knowledge, there does not exist a clear statement of what those  properties are.

 In this paper, we consider the weaker condition that    $\Lop\Lo/N^\alpha$ has a positive definite limit with $\alpha\in (0,1]$. Since it is the strength of the loadings that is being weakened and positive definiteness of $\Fop\Fo/T$  is maintained throughout, we use the terminology of {\em weaker loadings}.
We  obtain two results for average errors. The first result, which concerns  the low rank component, is  $\frac{1}{NT} \sum_{i=1}^N \sum_{t=1}^T \| \tilde C_{it}-C_{it}^0\|^2=O_p(\frac{1}{N})+O_p(\frac{1}{T})$. This result is somewhat surprising as it  is the same as in the strong factor case. The second result pertains to the error rate in estimating the space spanned by the factors. This rate is of interest because it  determines whether $\tF$ can be treated as though it was known in factor augmented regressions. We  obtain $\frac{1}{T}\sum_{t=1}^T \|\tilde F_t-\bm H^\prime F_t^0\|^2=O_p(\frac{1}{N^\alpha})+O_p((\frac{N^{1-\alpha}}{T})^2)$ which  is asymptotically $o_p(1)$ when $\alpha>0$ and $ \frac{N^{1-\alpha}}{T}\rightarrow 0$. This result implies an error rate  for the strong factor case of  $\alpha=1$ of  $\min(N,T^2)$, which is
better than the rate of $\min(N,T)$ previously derived. This improvement is made possible by a different proof technique that also leads to significant simplifications, hence of independent interest. The simplifications come partly from using higher level assumptions, and partly from using approximations to the original rotation matrix  $\bm H$ which also   make it possible to conduct inference using a  representation of the asymptotic variance that the user deems   most convenient.

Our main result is that  while  the strong factor assumption of $\alpha=1$ yields the fastest convergence rates  possible, and the estimates are inconsistent in the other extreme when $\alpha=0$, the principal component estimator for $\bm \Lambda$ and $\bm F$ continues to be consistent when $\alpha\in(0,1]$. In other words, except in the special case considered in \citet{onatski-joe:12}, the PC estimates are consistent.  We find that   asymptotic normality of  $\sqrt{N^\alpha}(\tilde F_t-\bm H^\prime F_t^0)$,  $\sqrt{T}(\tilde\Lambda_i-\bm H^{-1} \Lambda_i^0)$,  and $\min(\sqrt{N^\alpha},\sqrt{T})(\tilde C_{it}-C_{it}^0)$ do  require $\alpha > 1/2$ along with  some additional assumptions on $N$ and $T$, though $N$ is not required to grow at the same rate as  $T$. However, $\alpha>0$ suffices for consistency of the individual loadings $\tilde\Lambda_i$, while  $\alpha>1/3$  suffices for consistency of the individual factor estimates $\tilde F_t$.   Thus consistent estimates can be obtained with  weaker loadings than asymptotic normality.

It is natural to ask what happens when the loadings have  varying strength. That is, instead of a constant $\alpha$, we now have  $1 \ge \alpha_1\ge  \alpha_2\ge \ldots\ge  \alpha_r >0$.
 We show  that in this case,  what matters is the weakest loading, $\alpha_r$.  Asymptotic normality now requires  $\alpha_r>1/2$ but consistency of the individual estimates is possible without this requirement. Though the results are   in  agreement with the  constant $\alpha$  case,
setting up the framework  is  not so trivial  as it requires using  different normalization rates to study convergence of $\tF$ to $\bm F^0\bm H$ while allowing  the rotation matrix $\bm H$ to be consistent with the data generating process. The framework is more general than  that of \citet{freyaldenhoven:22} or \citet{uematsu-yamagata:est} which require specific assumptions on $\bm F^0 \bm H$ or $\bm H$, as discussed below.

The paper proceeds as follows. We start with the  simpler case   that $\alpha$ is the same for all factors and provide   the complete distribution theory for $\tilde F_t$, $\tilde \Lambda_i$ and $\tilde C_{it}$.  We then consider  the general case when $\alpha$ varies.
 Section 2 sets up the econometric framework and presents three useful preliminary results. Section 3 studies consistent estimation of the factors, the loadings, and introduces four asymptotically equivalent rotation matrices. The distribution theory is given in Section 4. Implications of weaker loadings for factor augmented regressions are discussed.  Section 5 studies the case of heterogeneous $\alpha$.

Throughout, matrices are written in bold-face to distinguish them from vectors.  As a matter of notation,  $\|\bm A\|^2=\sum_{i=1}^m\sum_{j=1}^n  |A_{ij}|^2=\text{Tr}(\bm A\bm A')$ is the squared Frobenius norm of a $m\times n$ matrix $\bm A$,
$\|\bm A\|^2_{sp}=\rho_{\max}(\bm A'\bm A)$ denotes the squared spectral norm of $\bm A$, where
$\rho_{\max}(\bm B)$ denotes  the largest eigenvalue of a positive semi-definite matrix $\bm B$.  Note that $\|\bm A\|_{sp}\le \|\bm A\| \le \sqrt{q} \|\bm A\|_{sp}$, where $q=rank(\bm A)$. Thus when the rank $q$ is fixed, the two norms are equivalent in terms of asymptotic behavior.

 \section{The Econometric Setup}

  We use  $i=1,\ldots N$ to index cross-section units and $t=1,\ldots T$ to index time series observations. Let  $X_i=(X_{i1},\ldots X_{iT})^\prime$ be a $T\times 1$ vector of random variables and
 $\bm X=(X_1,X_2,\ldots, X_N)$ be a $T\times N$  matrix.   The normalized data $\bm Z=\frac{\bm X}{\sqrt{NT}}$  admit
  singular value decomposition (\textsc{svd})
\[ \bm Z=\frac{\bm X}{\sqrt{NT}}
 =\bm U_{NT}\bm  D_{NT}\bm V_{NT}\pr=\bm U_{NT,k} \bm D_{NT,k} \bm V_{NT,k}\pr+ \bm U_{NT,N-k} \bm  D_{NT,N-k} \bm V_{NT,N-k}\pr\]
where $\bm U_{NT}'\bm U_{NT}=\bm I_T$ and $\bm V_{NT}'\bm V_{NT}=\bm I_N$.
In the above, $\bm  D_{NT,k}$ is a diagonal matrix of $k$ singular values $d_{NT,1},\ldots, d_{NT,k}$ arranged in descending order, $\bm U_{NT,k}, \bm V_{NT,k}$ are the corresponding left and right singular vectors respectively.   By the   \citet{eckart-young} theorem,    the best  rank $k$ approximation of $\bm Z$  is $\bm U_{NT,k}\bm D_{NT,k}\bm V_{NT,k}\pr$. This is obtained without imposing probabilistic assumptions on the data.

 We represent the data using  a static factor model with $r$ factors.
 In matrix form,
    \begin{eqnarray}
    \label{eq:dgp}
    \bm X&=&\bm F\bm \Lambda^\prime  + \bm e.
    \end{eqnarray}
To simplify notation, the subscripts indicating that $\bm F$ is $T\times r$ and $\bm \Lambda$ is $N\times r$ will be suppressed when the context is clear.  The common component $\bm C=\bm F\bm \Lambda\pr$ has reduced rank $r$ because $\bm F$ and $\bm \Lambda$ both have rank $r$.   The $N\times N$ covariance matrix of $\bm X$ takes the form
\[  \bm\Sigma_X=\bm \Lambda\bm\Sigma_F \bm\Lambda^\prime + \bm\Sigma_e=\bm \Sigma_C+\bm \Sigma_e .\]

A  strict factor model obtains  when the errors $e_{it}$ are cross-sectionally and serially uncorrelated so that $\bm\Sigma_e$ is a diagonal matrix. The classical factor model studied in \citet{anderson-rubin} uses the stronger assumption  that $e_{it}$ is iid and normally distributed. For economic analysis, this error structure is overly restrictive.  We work with the   approximate factor model formulated in \citet{chamberlain-rothschild} which allows the idiosyncratic errors to  be weakly correlated in both the cross-section and time series dimensions. In such a case,  $\bm \Sigma_e$ need not be a diagonal matrix.

Let $\bm F^0$ and $\bm \Lambda^0$ be the true values of $\bm F$ and $\bm \Lambda$. The model for  unit $i$ at time $t$ as
\[ x_{it}=\Lambda_i^{0\prime} F^0_t+e_{it}.\]
Letting  $e_i^\prime=(e_{i1},e_{i2},...,e_{iT})$ and $e_t^\prime=(e_{1t},e_{2t},...,e_{Nt})$, the model for unit $i$ is
\begin{eqnarray*}
 X_i&=&\bm F^0\Lambda^{0^\prime}_i+e_i
\end{eqnarray*}

Estimation of $\bm F^0$ and $\bm \Lambda^0$ in an approximate factor model with $r$ factors proceeds by minimizing  the sum of squared residuals:
 \begin{eqnarray*} \min_{\bm F,\bm \Lambda} \textsc{ssr}(\bm F,\bm \Lambda;r)&=&\min_{\bm F,\bm\Lambda} \frac{1}{NT} \|\bm X-\bm F\bm \Lambda^\prime\|^2\\ &=&\min_{\bm F,\bm\Lambda}\frac{1}{NT}\sum_{i=1}^N \sum_{t=1}^T (x_{it}-\Lambda_i^\prime F_t)^2.\end{eqnarray*}
As  $\bm F$ and $\bm \Lambda$ are not separately identified, we impose  the normalization restrictions
\begin{equation}
 \frac{\bm F'\bm F}T=I_r, \quad  \quad  \bm \Lambda'\bm\Lambda \quad \text{is diagonal} .
\label{eq:normalization1}
\end{equation}
The  solution  is the (static)  PC estimator defined as:
\begin{equation}
  (\tilde{\bm  F},\tilde{\bm \Lambda})=(\sqrt{T} \bm U_{NT,r},\sqrt{N}\bm V_{NT,r}\bm D_{NT,r}).
\end{equation}
 PC estimation of large dimensional approximate factor models must overcome two  challenges not present in the classical factor analysis of \citet{anderson-rubin}. The first  pertains to the fact that the errors are now allowed to be cross-sectionally correlated. The second issue arises because the $T\times T$ covariance matrix of $\bm X$ and the $N\times N$ covariance of $\bm X'$ are of infinite dimensions when $N$ and $T$ are large.     To study the properties of the PC estimates,
 we use  $\bm X=\bm F^0\bm \Lambda^{0\prime}+\bm e$ to obtain:
 \begin{eqnarray}
 \frac 1 {NT} \bm X \bm X'& = & \frac{\bm F^0(\bm\Lambda^{0\prime}\bm \Lambda^0)}{N}\frac{\bm F^{0'} }{T}+\frac{\bm F^0\bm \Lop \bm e^{\prime}}{N T} +
\frac{\bm e\bm \Lambda^0\bm  F^{0^\prime}}{N T}
+ \frac{\bm e \bm e^{\prime}}{N T}.
\label{identity0}
\end{eqnarray}
But   $\frac{1}{NT} \bm X\bm X^\prime=\bm U_{NT}\bm D_{NT}^2\bm U_{NT}^\prime$ and thus
 $ \frac{1}{NT}  \bm X\bm X' \tF =\tF \bm D_{NT,r}^2$. It follows that
\begin{eqnarray}
\frac{\bm F^0(\bm\Lambda^{0\prime}\bm \Lambda^0)}{N}\frac{\bm F^{0'}\tF }{T}+\frac{\bm F^0\bm \Lop \bm e^{\prime} \tF}{NT} +
\frac{\bm e\bm \Lambda^0\bm  F^{0^\prime}\tF}{NT}
+ \frac{\bm e \bm e^{\prime}\tF}{NT} &=&\tF \bm D_{NT,r}^2.
\label{eq:PC-F}
\end{eqnarray}
Rearranging terms  yields
\begin{eqnarray}
\tilde  F_t-\tilde {\bm H}_{NT,0}^\prime F^0_t&=&
\tilde {\bm D}_{NT} ^{-2} \bigg(\frac{1}{T}\sum_{t=1}^T \tilde  F_s \gamma_{st}+
\frac{1}{T}\sum_{s=1}^T \tilde F_s \zeta_{st}+
\frac{1}{T} \sum_{s=1}^T \tilde  F_s \eta_{st} + \frac{1}{T}
\sum_{s=1}^T \tilde  F_s \xi_{st}\bigg)
\label{eq:fourterms}
\end{eqnarray}
where
\begin{equation}
\label{eq:H0}
\bm H_{NT,0}=\bigg(\frac{\bm \Lambda^{0^\prime}\bm \Lambda^0}{N}\bigg)\bigg(\frac{\bm F^{0^\prime}\tF}{T}\bigg) \bm D_{NT,r}^{-2}
\end{equation}
$  \gamma_{st}=E(\frac{1}{N}e_s^\prime e_t)=E(\frac{1}{N} \sum_{i=1}^N e_{is} e_{it})$,
$\zeta_{st}=\frac{1}{N}e_s^\prime e_t-\gamma_{st}$,
$\eta_{st}= \bm F_s^{0^T}\bm \Lambda^{0^T} e_t/N$, and
$\xi_{st}= \bm F_t^{0^T} \bm \Lambda^{0^T} e_s/N$.
\citet{stock-watson-jasa:02,baing-ecta:02,bai-ecta:03} established properties of the PC estimator by analyzing the four terms in (\ref{eq:fourterms}) under certain assumptions, and this is by and large the approach that the literature has taken. We  work directly with the matrix norms of the terms in  (\ref{eq:PC-F}). This  makes it possible to obtain   simpler proofs   under more general assumptions.\footnote{An earlier version of the paper circulated as {\em Simpler proofs for approximate factor models of large dimensions} considers  $\alpha=1$ only.}

\subsection{Weaker Loadings: Homogeneous Case}

The defining characteristic of an approximate factor model is that the first $r$ largest population eigenvalues of   $\bm \Sigma_C$  diverge with $N$ while all remaining eigenvalues of $\bm \Sigma_C$ are zero, and all eigenvalues of $\bm \Sigma_e$ are bounded.  Previous works model  the `diverge with $N$' feature  by assuming that  $\bm F^{0^\prime} \bm F^0/T>0$ and $\bm \Lambda^{0^\prime} \bm \Lambda^0/N$ are positive definite in the limit.  These two conditions  have come to be known as the  strong factor structure.  \citet{onatski-joe:12} considers the other extreme that requires $\bm \Lambda^{0'}\bm \Lambda^0$ to have a positive limit  and shows that the factor estimates are inconsistent. There are many ways to accommodate weaker factor structures.    For example,  \citet{dgr-lasso} let  the  eigenvalues of $\bm \Sigma_e$ to be  large relative to those of $\bm \Sigma_C$. As discussed in \citet{onatski-joe:12}, such a setup  can be rewritten in terms of weaker loadings defined as $\bm \Lambda^{0'}\bm \Lambda^0/N^{\alpha}$ with $1\ge \alpha>0$, and is the approach that we will follow.

\paragraph{Assumption A1:}  Let $M<\infty$ not depending on $N$ and $T$ and define
\[ \delta_{NT}=\min(\sqrt{N},\sqrt{T}).\]

     \begin{itemize}
    \item[i]  Mean independence: $ E(e_{it}|\lio ,\fto)=0$.

\item[ii] Weak  (cross-sectional and serial) correlation in the errors.
\begin{itemize}
\item[(a)] $E  \Big[\frac 1 {\sqrt{N}} \sumiN [e_{it}e_{is}-E(e_{it}e_{is})]\Big]^2 \le  M $.
\item[(b)] For all $i$,  $\frac 1 T \sumtT\sumsT  | E (e_{it}e_{is})|  \le  M $. For all $t$, $\frac 1 N \sum_{i=1}^N\sum_{j=1}^N | E (e_{it}e_{jt})|\le M$.
\item[(c)] For all $t$, $\frac{1}{N\sqrt{T}} \|e_t^\prime \bm e'\|=\OpdNTione$ and for all $i$, $\frac{1}{T\sqrt{N}} \|e_i^\prime \bm e\|=\OpdNTione$.
\item [(d)]  $\|\bm e\|_{sp}^2=\rho_{\max}(\bm e'\bm e) =O_p(\max\{N,T\})$.
\end{itemize}
  \end{itemize}

  \paragraph{Assumption A2:}   (i) $E\|F_t^0||^4 \le M,$ $\plim_{T\rightarrow\infty} \frac{\Fop \Fo}{T}=\bm \Sigma_F>0$;\\
   (ii) $\|\Lambda_i^0\|\le M$, $\lim_{N\rightarrow\infty}\frac{\Lop\Lo}{N^\alpha}=\bm \Sigma_\Lambda>0$, for some $\alpha>0$ with $\alpha\in (0,1]$; \\
   (iii) the eigenvalues of $\bm\Sigma_\Lambda \bm \Sigma_F$ are distinct.

\paragraph{Assumption A3:}  For each $t$, (i) $E\|N^{-\a/2}\sum_i \Lambda^0_i e_{it}\|^2\le M$, (ii) $\frac{1}{NT} e_t^\prime\bm e^\prime\Fo=\OpdNTitwo$;  for each $i$, (iii) $E\|T^{-1/2} \sum_t F^0_t e_{it}||^2\le M$, (iv)
$\frac{1}{N^\alpha T}e_i'\bm e\Lo=O_p(\frac 1 {N^\alpha}) +O_p(\frac 1 {\sqrt{T N^\alpha}})$;
(v) $\Lop \bm e ' \Fo  = \sum_{i=1}^N\sum_{t=1}^T \Lambda_i^0 F_t^{0'} e_{it}  =O_p( \sqrt{N^\alpha T})$.

\paragraph{Assumption A4:} As $ N,T\rightarrow \infty$,  $(\frac N {N^\alpha}) \frac 1 T \rightarrow 0$ for the same  $\alpha$ in Assumption A2.

\vspace{0.05in}
Assumption A1.i uses  mean independence in place of moment conditions on $e_{it}$ as in previous work. Assumption A1.ii assumes weak time and cross-section dependence.       Assumption A1(d) is a bound on the maximum eigenvalues of $\bm e\pr\bm e$. For iid data with uniformly bounded fourth moments, the rate is implied by random matrix theory.   \citet{moon-weidner:17} extend the case to data that are weakly correlated across $i$ and $t$.  We use it  to obtain simpler proofs.  Assumption A2 implies  $\|\Fo\|^2/T=O_p(1)$  and $\|\Lo\|^2/N^\a=O_p(1)$. A2.(ii)   allows the $r$ eigenvalues of $\Lop\Lop$  to diverge at a rate of $N^\a$  with  $1\ge \alpha>0$.

Assumption A2 entertains weaker loadings by allowing $\bm \Lambda^{0'}\bm \Lambda^0/N^\alpha$ to have a positive definite limit with $1\ge \alpha>0$ which  nests the strong factor model as a special case. When $\alpha$ is  constant,  the strength of the loadings is homogeneous.  Note that the strength of the factor loadings affects the normalization of $\bm \Lambda^{0^\prime}\bm \Lambda^0$  but not $\bm F^{0^\prime}\bm F^0$.

  Assumptions A3.(ii) and (iv) reflect the more general setup that   $1\ge \alpha > 0$.     Parts of Assumption A3 also appear in \citet{baing-ecta:02}. When  the errors $e_{it}$ are independent,  Assumptions  A1 and A2 are enough to validate  A3.  The assumption should hold under weak cross-sectional and serial correlations.   Assumption A3  implies
\begin{eqnarray} \label{eq:FeeF}
 \frac {\Fop\bm e\bm e'\Fo}{NT}&=& \frac 1 N   \sum_{i=1}^N \Big[\bigg(\frac 1 {\sqrt{T}} \sum_t F^0_t e_{it}\bigg)\bigg(\frac 1 {\sqrt{T}} \sum_t F^0_t e_{it}\bigg)'\Big]
=O_p(1)\\
\label{eq:LeeL}
 \frac{\Lop\bm e'\bm e\Lo}{N^\a T}&=&\frac{1}{T} \sum_{t=1}^T \bigg[\bigg(\frac{1}{\sqrt{N^\a}} \sum_i \Lambda^0_i e_{it}\bigg)\bigg(\frac{1}{\sqrt{N^\a}} \sum_i\Lambda_i^0e_{it}\bigg)'\bigg]=O_p(1).
\end{eqnarray}


 Allowing for weaker factors comes at a cost. As stated in Assumption A4, which is new,   a small $\alpha$ must be compensated by a larger $T$. We assume that all variables in $\bm X$ are relevant in the sense of having a non-negligible common component. \citet{chao-swanson:0322,chao-swanson:0422} study selection of relevant variables in  factor augmented regressions.  To accommodate irrelevant variables, they assume $\frac{N}{N_1}\frac{1}{T}\rightarrow c$, $c>0$ and possibly $\infty$ where $N_1$ is the number of relevant variables. Assumption A4 rules this out.

The above assumptions are written  for analyzing weak loadings as defined in Assumption A2. The framework can be adapted to study  weak factors modeled as $\frac{\bm F ' \bm F}{T^\beta}$ being positive definite in the limit for any $1 \ge  \beta >0$. Whether we have weak loadings or weak factors,  the key feature is that  some  eigenvalues of $\Sigma_C$ will diverge at a rate   slower  than $N$.  Though the analysis proceeds as though the panel $X$ consists of data 
with i indexing units and t indexing time, the framework is also valid when  $i$ and $t$ take on other interpretation   provided that the data satisfy the assumptions above.

\subsection{Useful Identities and  Matrices}

   In the strong factor case,   $\bm D_{NT,r}^2$ is a diagonal  matrix of the $r$ largest  eigenvalues of $\bm Z'\bm Z=\frac{1}{NT}\bm X\bm X^\prime$ . The   singular values of $\bm Z$  are those of $\bm X$ divided by $\sqrt{NT}$. In practice, each column of $\bm Z$ is transformed to have unit variance so $d^2_{NT,j}$ is the fraction of variation  in $\bm Z$ explained by factor $j$.
 The following lemma shows that  to accommodate weaker factors, $\bm D^2_{NT,r}$ must  be scaled up by $N^{1-\alpha}$ to have  a limit  matrix $\bm D^2_r$ that is  full rank.

\begin{lemma}
\label{lem:Dr}
Let $\bm D_r^2$  be a diagonal matrix consisting of the  the ordered eigenvalues of $\bm \Sigma_\Lambda \bm \Sigma_F$. Under Assumption A, we have
\[  \Big(\frac N {N^{\alpha}} \Big) \bm D_{NT,r}^2 \pconv \bm D_r^2 >0, \quad 1\ge \alpha >0. \]
\end{lemma}

\paragraph {Proof:}
Proper normalization of $\bm D_{NT,r}^2$ is key to accommodating $1\ge \alpha>0$. The diagonal matrix $\frac{N}{N^\alpha}\bm D_{NT,r}^2$  consists of the $r$ largest eigenvalues of $\frac{1}{N^\alpha T} \bm X\bm X^\prime$, and
 \begin{eqnarray*}
 \frac 1 {N^{\alpha}T} \bm X \bm X'& = & \frac{\bm F^0(\bm\Lambda^{0\prime}\bm \Lambda^0)}{N^\alpha}\frac{\bm F^{0'} }{T}+\frac{\bm F^0\bm \Lop \bm e^{\prime}}{N^\alpha T} +
\frac{\bm e\bm \Lambda^0\bm  F^{0^\prime}}{N^\alpha T}
+ \frac{\bm e \bm e^{\prime}}{N^\alpha T}.
\end{eqnarray*}
Since
the largest eigenvalue of $\bm e\bm e'$ is of order $\max\{N,T \}$,  the largest eigenvalue of the last matrix is bounded
by $O_p(\frac N {N^\alpha} \frac 1 T )+ O_p(\frac 1 {N^{\alpha}}) \pconv 0$.   Furthermore,
$\|\bm e\|_{sp}=O_p(\sqrt{\max\{N,T\}})$, $\|\bm F^0\|_{sp}=O_p(T^{1/2})$ and $\|\bm \Lambda^0\|_{sp}=O_p(N^{\alpha/2})$.  A bound in  spectral norm for the second matrix on the right hand side is
\[\frac{\|\bm e\|_{sp}\|\bm F^0||_{sp}\|\bm \Lambda^0\|_{sp}}{T N^{\alpha}}
\le O_p\bigg(\sqrt{\frac N {N^\alpha} \frac 1 T} \bigg)+ O_p(\frac 1 {N^{\alpha/2}}) \pconv 0.\] The third matrix is the transpose of the second. Thus, the largest eigenvalues of the last three matrices converge to zero. By the matrix perturbation theorem, the $r$ largest eigenvalues of $\frac{1}{N^\alpha T} \bm X\bm X^\prime$ are determined by the first matrix on the right hand side. The eigenvalues of this matrix are the same  as those of
\[ \bigg( \frac{\bm\Lambda^{0\prime}\bm \Lambda^0}{N^\alpha}\bigg)\bigg(\frac{\bm F^{0'}\bm F^0 }{T}\bigg). \]
This matrix converges to $\bm \Sigma_\Lambda \bm \Sigma_F$  whose eigenvalues are $\bm D_r^2$, proving the lemma. $\Box$

Next, we turn to two matrices that will play important roles subsequently.  The first is the   matrix $\tilde {\bm  F^\prime }\bm F^0/T$. To obtain its  limit,  we multiply $(N/N^\alpha) \bm \tF'$ on each side of (\ref{eq:PC-F}) and use the fact that
$\bm \tF'\bm \tF=T$ to obtain
\begin{eqnarray} \label{eq:identity1}
 \Big(\frac{\bm \tF' \bm F^0}{T}\Big)\frac {\bm\Lambda^{0\prime}\bm \Lambda^0} {N^\alpha} \Big(\frac{\bm F^{0'} \bm \tF }{T}\Big)+\frac{\bm \tF' \bm F^0\bm \Lop \bm e^{\prime} \tF}{N^\alpha T^2} +
\frac{\bm\tF' \bm e\bm \Lambda^0\bm  F^{0^\prime}\tF}{N^\alpha T^2}
+ \frac{\bm \tF'\bm e \bm e^{\prime}\tF}{N^\alpha T^2} &=&  \frac N {N^\alpha}  \bm D_{NT,r}^2 .
\end{eqnarray}
The right hand side converges to a positive definite matrix (thus invertible) by Lemma \ref{lem:Dr}. The last three matrices on the left hand side converges in
probability to zero.  In particular,
\[ \|\frac{\bm \tF' \bm F^0\bm \Lop \bm e^{\prime} \tF}{N^\alpha T^2}\|\le \|\frac{\bm \tF' \bm F^0} T\| \|\bm \Lop \bm e^{\prime}\| \|\tF\|
\frac 1 {N^\alpha T} =O_p(\frac 1 {N^{\alpha/2}}) =o_p(1)  \]
and
\begin{equation}\label{eq:FeeF-hat} \|\frac{\bm \tF'\bm e \bm e^{\prime}\tF}{N^\alpha T^2}\| \le \rho_{\max}( \bm e \bm e')\frac {\|\tF\|^2} T  \frac 1 {N^\alpha T}
\le  \frac{ \max\{N,T \} } {N^\alpha T} O_p(1) =o_p(1). \end{equation}
The limit on the left hand side is thus determined by the first matrix, ie.
\begin{eqnarray} \label{eq:identity2}
 \Big(\frac{\bm \tF' \bm F^0}{T}\Big)\frac {\bm\Lambda^{0\prime}\bm \Lambda^0} {N^\alpha} \Big(\frac{\bm F^{0'} \bm \tF }{T}\Big)+o_p(1) =  \frac N {N^\alpha}  \bm D_{NT,r}^2.
\end{eqnarray}
  The limit of  $\tF'\Fop/T $ can be obtained from this representation.
\begin{lemma} Under Assumption A,
\begin{itemize}
\item[i.]  $\tF'\Fop/T \pconv \bm Q:=\bm D_{r} \bm \Upsilon' \bm \Sigma_{\Lambda}^{-1/2}$, where
 $\bm \Upsilon$ consists of the eigenvectors of the matrix  $\bm \Sigma_F^{1/2}\bm \Sigma_\Lambda \bm \Sigma_F^{1/2}$ with $\bm \Upsilon ' \bm \Upsilon =I_r$.
\item[ii.] For $\bm H_{NT,0}=\bigg(\frac{\bm \Lambda^{0^\prime}\bm \Lambda^0}{N}\bigg)\bigg(\frac{\bm F^{0^\prime}\tF}{T}\bigg) \bm D_{NT,r}^{-2},$ we have
\label{lemma:H}
$\bm H_{NT,0}\pconv \bm Q^{-1}$.
\end{itemize}
\end{lemma}
Part (i) is obtained by  taking limit on each side of (\ref{eq:identity2})   to yield
$ \bm Q \bm \Sigma_\Lambda \bm Q' = \bm D_r^2 $.  Since $\bm D_r^2$ is a positive definite matrix, it follows that $\bm Q $ is invertible.  Matrix $\bm Q$ can be expressed as  $\bm Q =\bm D_{r} \bm \Upsilon' \bm \Sigma_{\Lambda}^{-1/2}$
where $\bm \Upsilon$ consists of the orthonormal eigenvectors of the matrix  $\bm \Sigma_F^{1/2}\bm \Sigma_\Lambda \bm \Sigma_F^{1/2}$ such that $\bm \Upsilon' \bm \Upsilon=I_r$ (Bai, 2003).  Note that $\bm Q$ is unique up to a column sign change, just like $\tF$ is determined up to a column sign change.

 The rotation matrix $\bm H_{NT,0}$, first derived in \citet{stock-watson-di-wp}, has been used  to evaluate the precision of $\tilde {\bm F}$.
\citet{bai-ecta:03} shows that $\bm H_{NT,0}\pconv \bm Q^{-1}$ when $\alpha=1$. To accommodate weaker loadings, we consider
\[\bm H_{NT,0}=\bigg(\frac{\bm \Lambda^{0^\prime}\bm \Lambda^0}{N^{\alpha} }\bigg)\bigg(\frac{\bm F^{0^\prime}\tF}{T}\bigg) \Big(\frac N {N^\alpha} \bm D_{NT,r}^2\Big)^{-1}.\]
By  assumption, the first matrix on the right hand side is invertible while  the last two matrices are invertible by the previous lemmas.  Hence  $\bm H_{NT,0} \pconv \bm \Sigma_\Lambda \bm Q' \bm D_r^{-2}\equiv   \bm Q^{-1}$.  The matrix $\bm Q$ and its relation to $\bm H_{NT,0}$  are fundamental to the asymptotic theory in the strong factor case. Lemma \ref{lemma:H} shows that the relations  are  unaffected when  weaker loadings are allowed.

\section{Average Errors in Estimating the Factor Space}
This section has three parts. Subsection 1 presents results for consistent estimation of the space spanned by the factors.  Subsection 2 introduces four new rotation matrices. Subsection 3 uses these new matrices to show consistent estimation of the spanned by the loadings.

\subsection{The Factors}
 To establish consistent estimation of  $\tilde {\bm F}$ for  $\bm F^0$ up to rotation by $\bm H_{NT,0}$,  we multiply $\bm D_{NT,r}^{-2}$ to both sides of (\ref{eq:PC-F}) and use the definition of $\bm H_{NT,0}$ to obtain
\begin{eqnarray}
   \tF -\Fo \bm H_{NT,0}&=&\Big(\frac{\Fo\Lop\bm e^{\prime} \tF}{NT} +\frac{\bm e \Lo \Fop\tF}{NT}+ \frac{\bm e \bm e^{\prime}\tF}{NT} \Big) \bm D_{NT,r}^{-2} \nonumber\\
   &=&\Big(\frac{\Fo\Lop\bm e^{\prime} \tF}{N^\alpha T} +\frac{\bm e \Lo \Fop\tF}{N^\alpha T}+ \frac{\bm e \bm e^{\prime}\tF}{N^\alpha T} \Big)  \Big( \frac N {N^\alpha} \bm  D_{NT,r}^{2}\Big)^{-1}.
   \label{eq:factorspace}
\end{eqnarray}
 This implies
\begin{eqnarray*}
\frac 1 {\sqrt{T}}  \|\tF -\Fo\bm H_{NT,0}\| &\le& \left \{ 2 \Big(\frac {\|\Fo\| \|\tF\|}{T} \Big) \Big(  \frac 1 { \sqrt{T} N^{\alpha} }  \| \Lop\bm e'\| \Big)  +  \frac {\|\bm e \bm e^\prime \tF\|} {N^\alpha T^{3/2} }  \right \} \|\Big( \frac N {N^\alpha} \bm  D_{NT,r}^{2}\Big)^{-1}\|\\
& = & O_p\Big(  \frac 1 { \sqrt{T} N^{\alpha} }  \| \Lop\bm e'\| \Big)
+O_p\Big(\frac {\|\bm e \bm e^\prime  \bm \tF \|} {N^\alpha T^{3/2} }\Big).
\end{eqnarray*}
But  $ \frac 1 {\sqrt{T} N^\alpha }   \| \Lop\bm e'\| =O_p(\frac 1 {\sqrt{N^\alpha}}) $ by  (\ref{eq:LeeL}) and
$ \frac {\|\bm e \bm e^\prime  \bm \tF \|} {N^\alpha T^{3/2} }\le  \frac{\rho_{\max}(\bm e \bm e^\prime) \|\bm \tF\|}{N^\alpha T^{3/2} }= O_p(\frac 1 {N^{\alpha}}) + \frac N {N^{\alpha}} \frac 1 T O_p(1)$. Thus
\[ \frac 1 {\sqrt{T}}  \|\tF -\Fo\bm H_{NT,0}\|=O_p(\frac 1 {\sqrt{N^\alpha}})+\frac 1 { T}  \frac N {N^\alpha} O_p(1). \]
Squaring it gives the following proposition.

\begin{proposition}
 \label{prop:prop1}
Under Assumption A, the following holds:
\begin{eqnarray*}
\frac{1}{T} \|\tF -\Fo\bm H_{NT,0}\|^2=\frac{1}{T} \sum_{t=1}^T\|\tilde F_t-\bm H_{NT,0}^{\prime}F_t^0\|^2
=O_p\Big(\frac 1 {N^{\alpha}}\Big)  + \frac 1 {T^2} \bigg(\frac N {N^{\alpha}}\bigg)^2 O_p(1).
\end{eqnarray*}
\end{proposition}
The result is stated  in terms of squared Frobenius norm.  The  average error in estimating $\bm F$ vanishes at rate $O_p(\frac{1}{N^\alpha})+\frac{N^{2(1-\alpha)}}{T^2}O_p(1)$.
For $\alpha=1$,  Theorem 1 of \citet{baing-ecta:02} gives a convergence rate for the same quantity of $O_p(\frac{1}{N})+\frac{1}{T}O_p(1)$.
  The proposition here uses a different proof to obtain  a faster convergence rate of $O_p(\frac 1 N) +\frac 1 {T^2} O_p(1) $ for the strong factor case of $\alpha=1$. Implications of the proposition will be discussed subsequently.

\subsection{Equivalent Rotation Matrices}
 The rotation  matrix $\bm H_{NT,0}$ is a product of three $r\times r$ matrices and it is not easy to interpret. However,  we can rewrite  (\ref{eq:identity1}) as
\begin{equation}\label{eq:identity3}
 \Big(\frac{\bm \tF' \bm F^0}{T}\Big) \bm H_{NT,0}  =  \bm I_r-
 \left\{ \frac{\bm \tF' \bm F^0\bm \Lop \bm e^{\prime} \tF}{N^\alpha T^2} +
\frac{\bm\tF' \bm e\bm \Lambda^0\bm  F^{0^\prime}\tF}{N^\alpha T^2}
+ \frac{\bm \tF'\bm e \bm e^{\prime}\tF}{N^\alpha T^2} \right\} \Big(\frac N {N^\alpha}  \bm D_{NT,r}^2\Big)^{-1}.
\end{equation}
As $\bm\tFp \bm \Fo/T=O_p(1)$,  the  product
of $\frac {\tFp \Fo } T$ and $\bm H_{NT,0}$ is an identity matrix up to an negligible term if it can be shown that
 the three terms inside the bracket are small.  The next Lemma formalizes this result and shows that it also holds for four other rotation.

\begin{lemma}
 Under Assumption A,
\begin{itemize}
\item[i.] $
  \bm H_{NT,0} =\Big(\frac {\tFp \Fo } T \Big)^{-1}+  O_p\bigg(\frac 1 {\sqrt{N^\alpha T}}\bigg)+ \bigg[ O_p\bigg(\frac 1 {N^\alpha}\bigg)
 +\bigg(\frac N {N^{\alpha}}\bigg) \frac 1 T  O_p(1)\bigg].$

\item[ii.] For  $\ell=1,2,3,4$,
$ \bm H_{NT,\ell}=\bm H_{NT,0} +  O_p(\frac 1 {\sqrt{N^\alpha T}})+  O_p(\frac 1 {N^\alpha})
 +(\frac N {N^{\alpha}}) \frac 1 T  O_p(1)$,
where \newline $\bm   H_{NT,1}=(\Lp\bm \Lambda^0)(\tL\pr \bm \Lambda^0)^{-1}$, \newline
$   \bm H_{NT,2} =  (\Fp \bm F^0)^{-1}(\Fp \tF)$, \newline
$   \bm H_{NT,3} = (\tF\pr \bm F^0)^{-1}(\tF\pr \tF)=(\tF\pr \bm  F^0/T)^{-1}$, and  \newline
$\bm   H_{NT,4} =(\Lp \tL) (\tL^\prime\tL)^{-1} = (\Lp\tL/N) \bm D_{NT,r}^{-2}$.
\item[iii.]  $\bm H_{NT,\ell}\pconv  \bm Q^{-1}$.
\end{itemize}
\label{lem:lemmaH}
\end{lemma}

 Part (i),  shown in the Appendix, establishes the error  in approximating $\bm H_{NT,0}$ by $(\frac{\tFp \Fo}{T})^{-1}$  while part (ii) considers four additional approximations that provide  an intuitive interpretation of $\bm H_{NT} F^0_t$.  For example, $\bm H_{NT,2}$ is the coefficient matrix from projecting $\tilde{\bm  F}$ on the space spanned by $\Fo$ and $\bm H^\prime_{NT,2}F^0_t$ is asymptotically the fit from the projection.
These alternative rotation matrices were  used  in \citet{baing-joe:19} for $\alpha=1$. The above Lemma shows that they can still be used in place of $\bm H_{NT,0}$ when $\alpha<1$, but the adequacy of approximation will depend on $\alpha$.  Lemma \ref{lem:lemmaH} allows for simpler  proofs and  helps to interpret the error in estimating $\tF_t$ and $\tL_i$. But for consistency proofs, the result $\bm H_{NT,\ell}\bm H_{NT,0}^{-1}=I_r+o_p(1)$ often suffices, and it is implied by Lemma \ref{lem:lemmaH}.


\subsection{The Loadings and the Common Component}

The PC estimator satisfies
$  \frac 1 N \tLp \tL = \bm D_{NT,r}^2 $ and
we already have
$ \frac 1 {N^\alpha} \tLp \tL = \frac N {N^\alpha}\bm D_{NT,r}^2 \pconv \bm D_r^2. $
We can now provide a simple  consistency proof for $\tL$. Multiply $\frac 1 T \tF'$ to both sides of
$\bm X =\Fp\Lop +\bm e$ to obtain $ \frac 1 T \tFp\bm X =(\tFp \Fo/T) \Lop + \tFp\bm e /T$. We have
\begin{eqnarray}
 \tLp &=& \bm H_{NT,3}^{-1} \Lop + \tFp\bm e /T \nonumber \\
\tLp  - \bm H_{NT,3}^{-1} \Lop &=& \bm H_{NT,0}' \Fop \bm e/T + (\tF- \Fo \bm H_{NT,0})' \bm e /T \label{eq:LH3}
\end{eqnarray}
and thus
\[ \frac 1 {\sqrt{N}} \|\tLp - \bm H_{NT,3}^{-1} \Lop\|\le  \|H_{NT,0}\| \frac{\| \Fop \bm e\|}{T\sqrt{N}} +
\frac{ \|(\tF- \Fo \bm H_{NT,0})' \bm e\| } {T\sqrt{N} } \]
The first term   $\|\bm \Fo \bm e \|/(T \sqrt{N})=O_p(1/\sqrt{T})$ by equation (\ref{eq:FeeF}).  The second term is $O_p(\frac{1}{\sqrt{N^{1+\alpha}}})$, shown in the Appendix. Combining results and ignoring  terms  dominated by $O_p(T^{-1/2})$, we have
\[ \frac 1 {\sqrt{N}} \|\tLp - \bm H_{NT,3}^{-1} \Lop\| = O_p\bigg(\frac 1 {\sqrt{T}}\bigg)+ O_p\bigg(\frac 1 {\sqrt{N^{1+\alpha}}}\bigg).
\]

Squaring gives the next proposition:
\begin{proposition} Under Assumption A, the following holds
 \label{prop:prop2}
\begin{eqnarray*}
\frac 1 N \|\tL-\Lo (\bm H_{NT,0}')^{-1}\|^2=\frac{1}{N} \sumiN \|\tilde\Lambda_i- \bm H_{NT,0}^{-1}\Lambda_i^0\|^2=
O_p\bigg(\frac 1 T\bigg) + O_p\bigg(\frac 1 {N^{1+\alpha}}\bigg).
 \end{eqnarray*}
\end{proposition}
Note that  replacing $\bm H_{NT,3}$ by $ \bm H_{NT,0}$  does not affect the rate analysis.
In fact, we can use other rotation matrices to gain intuition. For example,  $\bm  H_{NT,1}^{-1}$ is obtained by regressing $\bm \tL$ on $\bm \Lambda_0$.  Hence $\bm \Lambda_0 (\bm H'_{NT,1})^{-1}$ is asymptotically the fit from projecting $\tilde{ \bm \Lambda}$ on the space spanned by $\bm \Lambda_0$, and $\tL-\Lo (\bm H_{NT,0}^\prime)^{-1}$ is the error from that projection.


While $\tF$ and $\tL$ only estimate $\bm F^0$ and $\bm \Lambda^0$ up to a rotation matrix, $\tilde{\bm C}$ does not depend on rotations and is directly comparable to $\bm C^0$.
\begin{proposition} Under Assumption A,
\begin{eqnarray*}
 \frac{1}{NT} \|\tilde{\bm C}-\bm C^0\|^2=\frac{1}{NT}\sum_{i=1}^N\sum_{t=1}^T \|\tilde C_{it}-C^0_{it}\|^2=\OpdNTitwo.
\end{eqnarray*}
  \label{prop:prop3}
\end{proposition}
\paragraph{Proof:} From
 $\tC-\bm C_0=\tF \bm \tLp-\bm F^0 \bm \Lambda^{0\prime}= (\tF-\bm F^0\bm H)\bm \tL'+ \bm F^0\bm H\bm \tL'-\bm F^0 \bm \Lambda^{0\prime}$, we have
\begin{eqnarray*}
\frac 1 { {NT}} \|\tilde{\bm C}-\bm C^0\|^2 & \le &  \frac {\|\tF-\bm F^0\bm H\|^2}{ {T}} \frac { \|\bm \tL'\|^2}{ {N}}+
\frac{ ||\bm F^0\bm H\|^2}{ {T}} \frac { \|\bm \tL'-\bm H^{-1} \bm \Lambda^{0\prime}\|^2}{ {N}} \\
&\le &  \Big[O_p\Big(\frac 1 {N^{\alpha}}\Big)  + \frac 1 {T^2} (\frac N {N^{\alpha}})^2 O_p(1)\Big] O_p\bigg(\frac {N^\alpha} N\bigg)
+ O_p\bigg(\frac 1 T\bigg) + O_p\bigg(\frac 1 {N^{1+\alpha}}\bigg)\\
 & = & O_p\bigg(\frac 1 N\bigg) + O_p\bigg(\frac 1 T\bigg) + \frac 1 {T^2} \frac N {N^\alpha} O_p(1)  =O_p(\delta_{NT}^{-2})
\end{eqnarray*}
where the second inequality follows from Propositions 1 and 2.
The term $\frac 1 {T^2} \frac N {N^\alpha} O_p(1)$ is dominated by $O_p(1/T)$ since  $\frac N {N^\alpha} \frac 1 T \rightarrow 0$ by Assumption A.4. Note this is the same rate as the strong factor case.
 $\Box$


\section{Distribution Theory}
\newcommand{\Na}{N^{\alpha}}

As we do not observe $\Fo$ or $\Lo$, we need an inferential theory for  $\tilde F_t$, $\tilde\Lambda_i$, and $\tilde C_{it}=\tilde F_t \tilde\Lambda_i^\prime$.  Theorems 1 and 2 of \citet{bai-ecta:03} establish that
\begin{subequations}
\begin{eqnarray}
\label{eq:FH0}
\sqrt{N}(\tilde F_t-\bm H_{NT,0}'F^0_t)&\dconv& \mathcal N(0, \bm D_{r}^{-2}\bm Q\bm \Gamma_t\bm Q'\bm D_{r}^{-2})\\
\sqrt{T}(\tilde \Lambda_i-\bm H_{NT,0}^{-1}\Lambda_i^0)&\dconv& \mathcal N(0,\bm Q^{'-1} \bm \Phi_i \bm Q^{-1}).
\label{eq:LH0}
\end{eqnarray}
\end{subequations}
by  positing appropriate central limit theorems (CLTs) appropriate for $\alpha=1$.  Assumption B accommodates weaker loadings.

\paragraph{Assumption B.}  The following holds  for each $i$ and $t$ as  $N, T\rightarrow \infty$:
\begin{eqnarray*}
   \frac{1}{\sqrt{N^\a}} \sum_{i=1}^N \Lambda^0 _i e_{it}\dconv \mathcal N(0,\bm \Gamma_t), \quad\quad
\frac{1}{\sqrt{T}}\sum_{t=1}^TF^0_te_{it}\dconv \mathcal N(0,\bm \Phi_i),
\end{eqnarray*}
 where $\Gamma_t$ and $\Phi_i$ are $r \times r$ positive definite matrices.

 The first CLT in Assumption B involves random variables over the cross section, while the second CLT involves random variables over
different time periods.  In view of the assumed weak dependence of $e_{it}$ over $i$ and $t$,  the two limiting distributions are independent and the convergence holds jointly.  The first CLT  uses a normalization of  $N^{\alpha/2}$ instead of the usual $N^{1/2}$ and is consistent with Assumption A2(ii). If each $e_{it}$ is iid with $Ee_{it}^2 =\sigma^2$, then the variance of the first term is $\sigma^2 \frac 1 {N^\alpha} \Lambda^{0\prime}\Lambda^0$, which converges to a positive definite matrix.  To see that CLT holds in spite of weaker loadings, suppose that $\Lambda_i^0 =\frac 1 {N^\tau} \delta_i$ where $\tau \in [0,1/2)$, and $\delta_i$ are either bounded constants or are  iid with $E(\delta_i\delta_i')>0$. Then for $\alpha=1-2\tau$, $N^{-\alpha/2} \sum_{i=1}^N \Lambda_i^0 e_{it} =N^{-1/2} \sum_{i=1}^N \delta_i e_{it}$, which is asymptotically normal by the standard arguments.

To  obtain  comparable asymptotic normality results when $1\ge \alpha> 0$ requires additional assumptions. To see why, first consider
 the limiting distribution $\tL$. From  (\ref{eq:LH3}),
\[ \sqrt{T}(\tilde \Lambda_i -\bm H_{NT,3}^{-1} \Lambda_i^0) = \bm H_{NT,3}' \frac{ 1}{\sqrt{ T}} \sum_{t=1}^T F_t^0 e_{it} + \frac{(\tF-\Fo\bm H_{NT,3})^\prime e_i}{\sqrt{T}}.\]
 The first term is asymptotically normal by Assumption B and $\bm H_{NT,3}^\prime\pconv \bm Q^{\prime -1}$ by Lemma \ref{lem:lemmaH}. For the second term, we show in Lemma {\ref{lem:lemma6} part (iii) that for each $i$,
\[  \frac 1 T e_i' (\tF- \Fo \bm H_{NT,\ell})   =O_p(\frac 1 {N^\alpha} ) + O_p(\frac 1 {\sqrt{T\Na}})+O_p(\frac N {N^{\alpha}T} ).\]
For $\sqrt{T}$ consistency of $\tL$, we need $\sqrt{T}$ times the above three terms to be negligible.
We thus require $\frac {\sqrt{T}} {N^\alpha} \rightarrow 0$ and $ \frac  1{\sqrt{T}} \frac N {N^\alpha} \rightarrow 0.$  A  necessary condition for both to hold  is  $\alpha>1/2$. These are listed as Assumption C(i)-(iii) below.
Note that the preceding equation  is $o_p(1)$ when $\alpha>0$. Thus if we merely consider consistency of $\tilde\Lambda_i$,  we only need $\alpha>0$.   It is only for $\sqrt{T}$ convergence and asymptotic normality that we need $\alpha>1/2$.

Similarly, for the  distribution of $\tilde F_t$,  we multiply  $ \tL (\tLp\tL)^{-1}$ to  both  sides of $\bm X= \Fo\bm \Lop +\bm e$ to obtain $\bm X \tL (\tLp\tL)^{-1}  =\Fo \bm \Lambda^{0'}\tL (\tLp\tL)^{-1}  + \bm e\tL (\tLp\tL)^{-1} $.  Using the definition of $\bm H_{NT,4}$,
\begin{eqnarray}
 \tF &=& \Fo \bm H_{NT,4}  +  \bm e\tL (\tLp\tL)^{-1}  \nonumber\\
&=& \Fo \bm H_{NT,4}  +  \bm e \Lo \bm H_{NT,4}^{\prime -1}(\tLp\tL)^{-1} + e(\tL-\Lo \bm H_{NT,4}^{\prime -1}) (\tLp\tL)^{-1}.\label{eq:FH4}
\end{eqnarray}
Thus
\begin{eqnarray*}
 \sqrt{N^\a} (\tilde F_t -\bm H_{NT,4}' F_t^0) &=&\bigg(\frac{\tL'\tL}{N^\a}\bigg)^{-1} \bm H_{NT,4}^{-1} \frac 1{\sqrt{N^\a}} \sum_{i=1}^N \Lambda_i^0 e_{it} +
\Big(\frac{\tLp\tL}{N^\alpha}\Big)^{-1} \frac{ (\tL-\Lo \bm H_{NT,4}^{\prime -1})'e_t }{\sqrt{N^\alpha}}
\end{eqnarray*}
 The first term is asymptotically normal by Assumption B.
 Now Lemma \ref{lem:lemma6} part (iv) shows that for each  $t$,  the average correlation between $e_t$ and the error from estimating $\tL$ is
\begin{equation}\label{eq:residual-F-estimation}  \frac 1 N e_t'(\tL-\bm \Lo \bm H_{NT,\ell}^{\prime -1})   = \frac 1 T \sqrt{\frac N {N^{\alpha}}}  O_p(1) +
\frac 1 {\sqrt{N^{1+\alpha}}} O_p(1)
+\frac 1 { T^{3/2}}  (\frac N {N^{\alpha}}) O_p(1) + \frac 1 {N^\alpha  \sqrt{T}} O_p(1). \end{equation}
In the strong factor case when $\alpha=1$, having the three terms vanish as $N,T\rightarrow \infty$ suffice for $\sqrt{N}$ consistency of $\tilde F_t$. With weaker loadings, we  can only get $\sqrt{N^\alpha}$ consistency of $\tilde F_t$, and this requires  $N/\sqrt{N^\alpha}$ times each of the four terms to be negligible. That is,
\[ \frac 1 T \frac{N^{3/2}}{N^\a} \rightarrow 0, \quad  N^{\frac 1 2 -\alpha}\rightarrow 0, \quad \frac 1 {T^{3/2}} \frac {N^2}{N^{3\alpha/2}}\rightarrow 0, \quad
\frac N {N^{3 \alpha/ 2}} \frac 1 {\sqrt{T}} \rightarrow 0.\]
The third and the fourth conditions hold if the first two along with C(iii) are satisfied. Thus, in addition to   $\alpha>1/2$, we also need  C(iv), $\frac{N}{T} N^{1/2-\alpha}\rightarrow 0$  to restrict the relation  between $N$ and $T$.
 But if instead of root-$N^{\alpha}$ asymptotic normality  we merely consider consistency of $\tilde F_t$,
 we only need $\alpha>1/3$. This follows upon multiplying   (\ref{eq:residual-F-estimation})  by $\frac N {N^{\alpha}}.$

We  collect the required conditions for $\sqrt{T}$ asymptotic normality of $\tL$ and $\sqrt{N^\alpha}$ asymptotic normality of $\tF$ into the following:
\paragraph{Assumption C:}
(i) $   \alpha >\frac 1 2$, (ii)  $  \frac {\sqrt{T}} {N^\alpha} \rightarrow 0$, (iii) $   \frac  1{\sqrt{T}} \frac N {N^\alpha} \rightarrow 0$, and (iv) $
\frac 1 T \frac{N^{3/2}}{N^\a} \rightarrow 0$.

\bigskip

 The above conditions reduce to $\sqrt{T}/N\rightarrow 0$ and $\sqrt{N}/T\rightarrow 0$ when $\alpha=1$. In general, they are distinctive restrictions, though some  conditions  may be redundant depending on the value of $\alpha$. Under Assumption C, the preceding analysis implies
\begin{subequations}
\begin{eqnarray}
\label{eq:F-H4}
 \sqrt{N^\a} (\tilde F_t -\bm H_{NT,4}' F_t^0) &=&\bigg(\frac{\tL'\tL}{N^\a}\bigg)^{-1} \bm H_{NT,4}^{-1} \frac 1{\sqrt{N^\a}} \sum_{i=1}^N \Lambda_i^0 e_{it} + o_p(1)\\
\label{eq:L-H3}
 \sqrt{T} (\tilde \Lambda_i -\bm H_{3,NT}^{-1} \Lambda_i^0) &=& \bm H_{NT,3}'\frac 1 {\sqrt{T}} \sum_{t=1}^T F_t^0 e_{it} + o_p(1).
\end{eqnarray}
\end{subequations}

\begin{proposition}
\label{prop:propFL}
Under Assumptions A, B, and C and the normalization that $\bm F'\bm F/T=I_r$ and $\bm\Lambda'\bm\Lambda$ is diagonal, we have,  as $N,T\rightarrow\infty$,
\begin{eqnarray*}
\sqrt{N^\a}(\tilde F_t-\bm H_{NT,4}'F^0_t)&\dconv& \mathcal N(0,  \bm D_{r}^{-2}\bm Q\bm \Gamma_t\bm Q'\bm D_{r}^{-2}),\\
\sqrt{T}(\tilde \Lambda_i-\bm H_{NT,3}^{-1}\Lambda_i^0)&\dconv& \mathcal N(0,\bm Q^{'-1} \bm \Phi_i \bm Q^{-1}).
\end{eqnarray*}
\end{proposition}
The Proposition establishes $(\sqrt{N^\alpha},\sqrt{T})$  asymptotic normality
of  $(\tilde F_t,\tilde\Lambda_i)$.    Though the conditions are more  restrictive than the strong factor case,  the results are  more general  than previously understood. Conditions  weaker than Assumption C  may be possible. For example, if $T/N\rightarrow c\in(0,\infty)$, then $\alpha>1/2$ is sufficient for asymptotic normality.

For hypothesis testing, there is no need to know $\alpha$. The limiting variance  $\frac{1}{N^\alpha} \bm D_r^{-2} \bm Q \bm \Gamma_t \bm Q^\prime \bm D_r^{-2}$  can be consistently estimated by
$\bm  D_{NT,r}^{-2}(\sum_{i=1}^N \tilde \Lambda_i \tilde \Lambda_i' \tilde e_{it}^2)  \bm  D_{NT,r}^{-2}$ if we assume no cross-section correlation in $e_{it}$. If cross-section and time dependence are allowed, the CS-HAC variance estimator developed in \citet{baing-ecta:06} can be used. The square root of the $k$-th diagonal element of this matrix gives the standard error of the $k$-th component of $\tilde F_t$. Similarly,
$\frac{1}{T} \bm Q^{\prime^{-1}}\bm \Phi_i \bm Q^{-1}$ can be consistently estimated by $\frac 1 T \sum_{s=1}^T \tilde F_s \tilde F_s' \tilde e_{is}^2$
(assuming no serial correlation in $e_{it}$).  These formulas are identical to the case of strong factors.



Although the limiting covariance matrices look different from those given in (\ref{eq:FH0}) and (\ref{eq:LH0}),  they are mathematically identical because of the different ways to represent $\bm H_{NT}$ and thus $\bm Q$. Indeed,
other representations of  the asymptotic variances can be used. For example, since
$ (\tLp\tL/N^\a)^{-1} \bm H_{NT,4}^{-1 } = (\Lp \tL /N^\a)^{-1} =\bm H_{1,NT}^{\prime} (\Lop\Lo/N^\a)^{-1}$ and  $\bm H_{NT,\ell} \pconv \bm Q^{-1}$ for all $H_{NT,\ell}$ considered in Lemma \ref{lem:lemmaH}, we also have
\begin{eqnarray}
  \sqrt{N^\a}(\tilde F_t -\bm H_{NT,4}' F_t^0) &=&\bm H_{NT,1}' \bigg(\frac {\Lop\Lo} {N^\a}\bigg)^{-1}  \frac{ 1}{\sqrt{ N^\a}} \sum_{i=1}^N \Lambda_i^0 e_{it} + o_p(1)
\label{eq:Fh4}\\
&\dconv&\mathcal N(0, \bm Q^{-1'} \bm\Sigma_\Lambda^{-1} \bm \Gamma_t \bm\Sigma_{\Lambda}^{-1} \bm Q)\nonumber.
\end{eqnarray}
 From  $\bm H_{NT,3}'=\bm H_{NT,2}^{-1} (\Fp \Fo/T)^{-1}$  and using (\ref{eq:L-H3}), it also holds that
\begin{eqnarray}
 \sqrt{T} (\tilde \Lambda_i -\bm H_{NT,3}^{-1} \Lambda_i^0) &=& \bm H_{NT,2}^{-1} \bigg(\frac{\Fp \Fo}{T}\bigg)^{-1}  \frac 1 {\sqrt{T}} \sum_{t=1}^T F_t^0 e_{it} + o_p(1) \label{eq:Lh3}\\
&\dconv & \mathcal N(0, \bm Q \bm\Sigma_F^{-1} \bm \Phi_i \bm\Sigma_F^{-1} \bm Q^{^\prime})\nonumber.
\end{eqnarray}
 As the factor estimates are asymptotically normal regardless of the choice of the rotation matrix,  one can use the  most convenient representation for inference.

 While there are many ways to represent the sampling error of $\tilde F_t$ and $\tilde \Lambda_i$,  the properties of $\tilde C_{it}$ are invariant to the choice of $\bm H_{NT,\ell}$, so we can simply write $\bm H_{NT}$. By definition, $C^0_{it}=\Lambda_i^{0\prime} F^0_t $ and $\tilde C_{it}=\tilde\Lambda_i^\prime \tilde F_t^0 $. Adding and subtracting terms
\begin{eqnarray*}
\tilde C_{it}-C_{it}^0&=& \Lambda_i^{0\prime}\bm H_{NT}^{'-1} (\tilde F_t-\bm  H_{NT}'F_t^0)' +(\tilde\Lambda_i-\bm H_{NT}^{-1}\Lambda_i^0)^\prime \tilde F_t\\
&=&\Lambda_i^{0\prime}\bm H_{NT}^{'-1} (\tilde F_t-\bm  H_{NT}'F_t^0)' +F_t^{0'} \bm H_{NT}(\tilde\Lambda_i-\bm H_{NT}^{-1}\Lambda_i^0)+O_p(\frac 1 {\sqrt{TN^\alpha}}).
\end{eqnarray*}
Using   (\ref{eq:Fh4}) and (\ref {eq:Lh3}), we have
\begin{eqnarray*}
(\tilde C_{it}-C_{it}^0)
&=&\frac{1}{\sqrt{N^\a}} \Lambda_i^{0\prime}\bigg(\frac{\Lop\Lo}{N^\a}\bigg)^{-1}  \frac 1{\sqrt{N^\a}} \sum_{i=1}^N \Lambda_i^0 e_{it} \\
&&+\frac{1}{\sqrt{T}} F_t^{0'} \bigg(\frac{\Fp \Fo}{T}\bigg)^{-1}  \frac 1 {\sqrt{T}} \sum_{t=1}^T F_t^0 e_{it} + o_p(N^{-\alpha/2})+o_p( T^{-1/2})
\end{eqnarray*}
 This leads to  the distribution theory for the estimated common components.
\begin{proposition}
\label{prop:prop4}
Under Assumptions A, B and C,
we have,  as $N,T\rightarrow\infty$,
\begin{eqnarray*}
\frac{\tilde C_{it}-C^0_{it}}{\sqrt{ \frac 1  {N^\a}   W^\Lambda_{NT,it}      +\frac{1}{T} W^F_{NT,it} }}&\dconv& \mathcal N(0,1)
\end{eqnarray*}
where $W^\Lambda_{it}=\Lambda_i^{0\prime} \bm \Sigma_\Lambda^{-1} \bm \Gamma_t \bm \Sigma_\Lambda^{-1} \Lambda_i^0$ and $W^F_{it}=F_t^{0\prime} \bm \Sigma_F^{-1}\bm \Phi_i \bm \Sigma_F^{-1} F_t^0$.
\end{proposition}

The proposition implies a convergence rate of $\min\{ \sqrt{N^{\alpha}}, \sqrt{T}\}$ for the estimated common components.
To estimate the standard error of $\tilde C_{it}$, there is also no need to know $\alpha$.
Assuming no cross-sectional correlation in $e_{it}$, the term $\frac 1  {N^\a}   W^\Lambda_{NT,it}$ can be consistently estimated by
$\tilde \Lambda_i^{\prime} (\sum_{i=1}^N \tilde \Lambda_i \tilde \Lambda_i')^{-1} (\sum_{i=1}^N \tilde \Lambda_i \tilde \Lambda_i' \tilde e_{it}^2) (\sum_{i=1}^N \tilde \Lambda_i \tilde \Lambda_i')^{-1} \tilde \Lambda_i$ . Furthermore, assuming no serial correlation in $e_{it}$,
 $\frac{1}{T} W^F_{NT,it}$ can be consistently estimated by
$\frac 1 T \tilde F_t' (\sum_{s=1}^T \tilde F_s \tilde F_s' \tilde e_{is}^2) \tilde F_t$. A variance estimator that allows $e_{it}$ to be correlated is discussed in \citet{baing-ecta:06}.

\newcommand{\eps}{\epsilon}

\subsection{Implications for Factor-Augmented Regressions}
The results have  implications for empirical work. Consider the infeasible regression
\begin{equation*}
y_{t+h}= \gamma^\prime F^0_t+\beta^\prime W_t+\epsilon_{t+h}\\
\end{equation*}
where $W_t$ is observed  but $F^0_t$ is not,  $\mathbb E(W_t\epsilon_{t+h})=0$ and $\mathbb E(F^0_t\epsilon_{t+h})=0$.
The feasible regression  upon replacing $F^0_t$ with  $\tF$ is
\begin{eqnarray}
\label{eq:favar}
y_{t+h}&=& \gamma^\prime \bm H_{NT}^{\prime -1} \tilde F_t+\beta^\prime W_t+ \epsilon_{t+h}+\gamma^\prime\bm H_{NT}^{\prime -1} (\bm H'_{NT}F_t^0-\tilde F_t).
\end{eqnarray}
\citet{baing-ecta:06} shows under the strong factor assumption $\tF$  can be used in a second step regression to obtain standard normal inference without the need for standard error adjustments if $\sqrt{T}/N\rightarrow 0$.
 To establish the comparable conditions   when $1 \ge  \alpha >0$, we need to analyze the correlation between the regressors $(\tilde F_t, W_t)$ and the errors $\epsilon_{t+h}$ as well as with the first step estimation error $(\bm H'_{NT}F_t^0-\tilde F_t)$.

\begin{lemma}
Let $\hat z_t=(\hat F_t^{\prime}, W_t^\prime)^\prime$  be used in place of $z_t=(F_t^\prime,W_t^\prime)^\prime$ in the factor-augmented regression (\ref{eq:favar}), and
 denote $\delta^0=(\gamma^\prime \mathbf H_{NT}^{\prime -1}   ,\beta^\prime)^\prime$.
Suppose $\bm W^\prime \bm e \bm\Lambda^0=\sum_{i=1}^N \sum_{t=1}^T W_t \Lambda_i^{0\prime}e_{it}=O_p(\sqrt{N^\alpha T})$  and  Assumptions A, B, and C hold. Then
\[  \sqrt{T}(\hat\delta-\delta^0)\dconv N(0,\mathbf J^{'^{-1}} \mathbf \Sigma_{zz}^{-1}\mathbf \Sigma_{zz,\epsilon}\mathbf \Sigma_{zz}^{-1}\mathbf J^{-1})\]
where $\mathbf J$ is the limit of $\mathbf J_{NT}=\text{diag}(\mathbf H_{NT}', \mathbf I_{\text{dim}(W)})$.
\label{lem:favar}
\end{lemma}
To understand the result, we need to show that the errors in (\ref{eq:favar}) are asymptotically uncorrelated with the regressors. Notice that
\[ \frac{1}{\sqrt{T}}\sum_{t=1}^T \tilde F_t  \epsilon_{t+h}  =  \mathbf H_{NT}' \frac{1}{\sqrt{T}} \sum_{t=1}^T F^0_t \epsilon_{t+h} +\frac{1}{\sqrt{T}}\sum_{t=1}^T (\tilde F _t-\mathbf H_{NT}' F_t^0) \epsilon_{t+h}. \]
We need to show that the second term is $o_p(1)$ and that $W_t$ is asymptotically uncorrelated with $(F_t^{0\prime}\mathbf H_{NT}- \tilde F_t^{\prime})$.   More precisely we need to show that each of  the following three terms is $o_p(1)$:
\[ (i),  \frac{1}{\sqrt{T}}\sum_{t=1}^T (\tilde F _t-\mathbf H_{NT}' F_t^0) \epsilon_{t+h}, \quad (ii),
  \frac{1}{\sqrt{T}}\sum_{t=1}^T  \tilde F_t(F_t^{0\prime}\mathbf H_{NT}- \tilde F_t^{\prime}), \quad (iii),
\frac{1}{\sqrt{T}}\sum_{t=1}^T  W_t (F_t^{0\prime}\mathbf H_{NT}- \tilde F_t^{\prime}).\]


Note first that $ \frac 1 {\sqrt{T}} \epsilon' (\tF- \Fo \bm H_{NT,\ell})  =o_p(1)$ upon replacing $e_i=(e_{i1},..., e_{iT})'$   by $\eps =(\eps_{1+h},...\eps_{T+h})$ in Lemma \ref{lem:lemma6}(iii). Furthermore, $ \frac 1 {\sqrt{T}}  \bm W' (\tF- \Fo \bm H_{NT,\ell})  =o_p(1)$  upon replacing  $\bm F^0$ by $\bm W=(W_1,\ldots, W_T)^\prime$ in Lemma \ref{lem:lemma6}(i). Hence all three terms are $o_p(1)$ under the assumptions of the analysis. As a consequence
 $\tF$ can be used in the augmented regression and yield standard normal inference as though it were $\bm F$, though the conditions $\alpha>1/2$ and $\sqrt{T}/N^\alpha\rightarrow 0$  are stronger  than for the $\alpha=1$ case. However,  if we only want consistent estimates instead of root-$T$ consistency and normality,   Assumption A.4 is sufficient; there is no need for Assumption C or  $\alpha>1/2$.


\section{Heterogeneous $\alpha$}

The foregoing analysis assumes that all loadings have the same strength as indicated by the constant $\alpha$.  In this section, we allow $\alpha$ to vary across factors. Let
$ 1\ge  \alpha_1 \ge \alpha_2 \ge \cdots \ge \alpha_r>0$ so that the weakest loading has strength $\alpha_r>0$. Now  define the $r\times r$ normalization matrix
\[ \bm \B=\diag(N^{\alpha_1/2},...,N^{\alpha_r/2}), \]
noting that $\|\B\|\le N^{\alpha_1/2}$ and $\|\Bi\|\le N^{-\alpha_r/2}$. Because of varying loadings strength, the bounds in Assumption A3 need to be replaced by the following:

\paragraph{Assumption A3':}  For each $t$, (i) $E\|\Bi\sum_i \Lambda^0_i e_{it}\|^2\le M$, (ii) $\frac{1}{NT} e_t^\prime\bm e^\prime\Fo=\OpdNTitwo$;  for each $i$, (iii) $E\|T^{-1/2} \sum_t F^0_t e_{it}||^2\le M$, (iv)
$\frac{1}{T}e_i'\bm e\Lo\Bi=O_p(\frac 1 { N{^{\alpha_r/2}}}) +O_p(\sqrt{\frac {N^{\alpha_1}}{T N^{\alpha_r}} })$;
(v) $\Lop \bm e ' \Fo  = \sum_{i=1}^N\sum_{t=1}^T \Lambda_i^0 F_t^{0'} e_{it}  =O_p( \sqrt{N^{\alpha_1} T})$.

\medskip

Changes to  Assumption A2 and A4 are also needed. We will  maintain  A2(i) and A2(iii), but
 A2.(ii) previously stated as $\frac{\bm \Lambda^{0'}\bm \Lambda^0}{N^\alpha}=\Sigma_{\Lambda}$ for $\alpha \in (0,1]$ is now replaced by
\paragraph{Assumption A2'.(ii):}
$ \B^{-1}\bm \Lop \bm \Lo \B^{-1} \pconv  \bm \Sigma_\Lambda>0$, where $\bm \Sigma_\Lambda$ is diagonal.

\vspace{0.2in}
Instead of  $\frac{N}{N^{\alpha}}\frac{1}T\rightarrow 0$, we need to replace Assumption A4 by
\paragraph{Assumption A4':} As $N, T\rightarrow \infty$, $\frac{N}{N^{\alpha_r}}\frac{1}{T}\rightarrow 0$ as $N,T\rightarrow \infty$.
\bigskip

We will need new identities to accommodate varying $\alpha$.  In  Lemma \ref{lem:Dr}, we see that the first $r$ largest eigenvalues of
$\frac{1}{N T} \bm X\bm X^\prime$ are determined by the matrix
$\frac1 {NT} \bm F^0(\bm\Lambda^{0\prime}\bm \Lambda^0)\bm F^{0'}$. A matrix with equivalent eigenvalues that we  now use is
 $\frac 1 N (\Bi\bm \Lop \bm \Lo \Bi)(\B\Fop \Fo/T) \B $. By assumption,
$(\Bi \bm \Lop \bm \Lo \Bi )$ and $\Fop \Fo/T$ both converge in probability to positive definite matrices. Furthermore,
the first $r$ eigenvalues is the order of $\B^{2}/N$. Since $\bm D_r^2$ is a positive definite matrix,  the comparable result to $\frac{N}{N^\alpha} \bm D_{NT,r}^2 \pconv \bm D_r^2>0$ in Lemma \ref{lem:Dr} is
\[ N \B^{-2} \bm D_{NT,r}^2 \pconv \bm D_r^2 >0.  \]

Next, we need to find a result  comparable to  positive definiteness of $\bm Q$ in Lemma \ref{lemma:H}, where $\bm Q$ is the limit of $\tF'\bm F^0/T$.
To proceed, we use
 $\frac{1}{NT} \bm X\bm X'\tF =\tF \bm D_{NT,r}^2$ and $\tF'\tF/T=\bm I_r$ to obtain
\begin{equation*} \label{eq:BFXXFB} \frac 1 {T^2} \Bi \tF' \bm X\bm X^\prime \tF\Bi = N \B^{-2} \bm D_{NT,r} ^2.
\end{equation*}
The leading  term of the left hand side  is
\[( \Bi \tF' \bm F^0\B/T)  (\Bi \bm\Lambda^{0\prime}\bm \Lambda^0 \Bi)  (\B \bm F^{0'} \tF \Bi /T) \pconv \bm D_r^2 \]
Since $(\Bi \bm\Lambda^{0\prime}\bm \Lambda^0 \Bi)\pconv \bm \Sigma_\Lambda>0$, and $\bm D_r^2 >0$,  the
 limit of $\Bi \tF' \bm F^0\B /T$ is invertible. We continue to denote its limit as $\bm Q$.

\subsection{Average Error in Estimation of the Factor Space}
Finding an appropriate  definition of the rotation matrix  to accommodate heterogeneous $\alpha$ is  delicate because the moments have to  be normalized in accordance with the strength of the loadings. We proceed by  right multiplying $\Bi$ to
$ \frac 1 {TN}  \bX \bX' \tF =  \tF  \DNTr^2 $ to get
\[ \frac 1 {TN}  \bX \bX' \tF\Bi =  \tF  \DNTr^2 \Bi \equiv \tF \B (\B^{-2} \DNTr^2). \]
Expanding $\bX \bX'$, and multiplying $N$ on each side gives
 \begin{eqnarray*}
  \bm F^0(\bm\Lambda^{0\prime}\bm \Lambda^0) \frac{\bm F^{0'}\tF \Bi }{T}+\frac{\bm F^0\bm \Lop \bm e^{\prime}\tF \Bi}{ T} +
\frac{\bm e\bm \Lambda^0\bm  F^{0^\prime}\tF \Bi}{T}
+ \frac{\bm e \bm e^{\prime}\tF \Bi}{ T} =  \tF \B (N \B^{-2} \DNTr^2).
\end{eqnarray*}
We  now define
\begin{equation}
\label{eq:Hbar}  \bar \bH_{NT} = \Big(\Bi (\bm\Lambda^{0\prime}\bm \Lambda^0) \frac{\bm F^{0'}\tF \Bi }{T}
+\frac{\Bi \Lop \bm e^{\prime}\tF \Bi}{T}  \Big) (N \B^{-2} \DNTr^2)^{-1}.
\end{equation}
Note that $\| \bar \bH_{NT}||=O_p(1)$. We obtain the following  average errors in estimating the  space spanned by the factors and the loadings.
\begin{proposition}
 Let  $\bar \bH_{NT}$ be defined as in (\ref{eq:Hbar}) and  $\bH_{NT}= \B \bar \bH_{NT} \Bi$. Then under Assumptions A',
\begin{itemize}
\item[i.] $ \frac 1 T \| (\tF -\Fo  \bH_{NT})\|^2 =O_p(\frac 1 {N^{\alpha_r}} ) + O_p( \frac {N^{2-2\alpha_r}} {T^2}  )$;
\item[ii.] $ \frac 1 N \|\tLp - \bm H_{NT,3}^{-1} \Lop\|^2 =  O_p(\frac {N^{\alpha_1-\alpha_r}} T )+\frac 1  {N^{(1+\alpha_r)}} {O_p(1)}.$
\end{itemize}
\label{prop:prop6}
\end{proposition}
When $\alpha$ is homogeneous,  $\bm B_N=\sqrt{N^\alpha} \bm I_r$ and $\bar \bH_{NT}$ is  $\bm H_{NT,0}=
 (\frac{\bm \Lambda^{0'}\bm \Lambda^0) }{N^\alpha}) $
$(\frac{\bm F^{0'}\tF }{T})\frac{N^{\alpha}}{N}\bm D_{NT,r}^{-2}$ defined earlier
since $ \frac{\bm \Lambda^{0'}\bm e' \tF }{NT}\bm  D_{NT,r}^{-2}$  is negligible.  With heterogeneous $\alpha$, the second term is still negligible, but including it in $\bar {\bm H}_{NT}$  makes it possible to use arguments that lead to  better convergence rates.   In particular,  the definition of $\bar \bH_{NT}$ implies:
\begin{eqnarray} \label{eq:FBFoB}\tF \B-\Fo \B \bar \bH_{NT} & = &\frac{\bm e\bm \Lambda^0  \bm  F^{0^\prime}\tF \Bi}{T}(N \B^{-2} \DNTr^2)^{-1}
+\frac{\bm e \bm e^{\prime}\tF \Bi}{ T} (N \B^{-2} \DNTr^2)^{-1} \nonumber\\
&=& a+b .
 \end{eqnarray}
The Appendix shows that $\|a\|=O_p(\sqrt{T})$ and $\|b\|=\frac{\max(N,T)} {\sqrt{T} N^{\alpha_r/2}}$, which together imply
\begin{equation} \label{FBFBH} \| \tF \B-\Fo \B \bar{\bm  H}_{NT}\|=O_p(\sqrt{T})+\frac { N^{1-\frac 1 2\alpha_r}} {\sqrt{T}}  O_p(1)  \end{equation}
Since $ \bH_{NT} \B= \B \bar \bH_{NT}$,  it follows that
\[ \| \tF -\Fo  \bH_{NT}\| =\| (\tF \B-\Fo \B \bar \bH_{NT})\Bi\|\le \| (\tF \B-\Fo \B \bar \bH_{NT})\| \|\Bi\|.  \]
Combining  (\ref{FBFBH}) with the fact that  $\|\Bi\|\le N^{-\alpha_r/2}$ yields
\[  \frac 1 {\sqrt{T}}\| (\tF -\Fo  \bH_{NT})\|
\le O_p(1)N^{-\alpha_r/2} +\frac { N^{1-\alpha_r}} T  O_p(1).  \]
Part (i) of the Proposition follows.
  A generalization of Lemma \ref{lem:favar} concerning the factor augmented regression is that  $\alpha_k>1/2$ and $\sqrt{T}/N^{\alpha_k}\rightarrow 0$ will be needed for standard normal inference if we use estimates of the largest $k$  factors   in two step regressions as though  $\bm F_1,\ldots, \bm F_k$ (where $k\le r$) were observable.

 For part (ii), we have  $\tLp = \bm H_{NT,3}^{-1} \Lop + \tFp\bm e /T $, where $ \bm H_{NT,3}^{-1}=(\tF\pr \bm  F^0/T)^{-1}$. Adding and subtracting terms
\begin{eqnarray*}
\tLp  - \bm H_{NT,3}^{-1} \Lop &=& \bm H_{NT}' \Fop \bm e/T + (\tF- \Fo \bm H_{NT})' \bm e /T \\
 &=& \bm H_{NT}' \Fop \bm e/T + \Bi(\tF\B- \Fo \bm H_{NT} \B )' \bm e /T.
\end{eqnarray*}
Taking norms,
\[
 \frac 1 {\sqrt{N}} \|\tLp - \bm H_{NT,3}^{-1} \Lop\|\le  \|\bH_{NT}\| \frac{\| \Fop \bm e\|}{T\sqrt{N}} +
\frac{ \|\Bi\| \|(\tF \B- \Fo \bm H_{NT} \B)' \bm e\| } {T\sqrt{N} }. \]
 Now
$ \|\bH_{NT}\| \frac{\| \Fop \bm e\|}{T\sqrt{N}}=N^{(\alpha_1-\alpha_r)/2} O_p(1/\sqrt{T})$ since
$\|\bH_{NT}\|\le \|\B\|\|\bar \bH_{NT}\| \Bi\|\le N^{(\alpha_1-\alpha_r)/2}O_p(1)$.  The Appendix shows that,  under$\frac N {TN^{\alpha_r}} \rightarrow 0$ as in Assumption A4',
\begin{eqnarray}
\frac{ \|\Bi\| \|(\tF \B- \Fo \bm H_{NT} \B)' \bm e\| } {T\sqrt{N} }
& =&\frac 1  {N^{(1+\alpha_r)/2}} {O_p(1)} + o_p(\frac 1 {\sqrt{T}}).
\label{eq:LHL}
\end{eqnarray}
 Thus
$ \frac 1 {\sqrt{N}} \|\tLp - \bm H_{NT,3}^{-1} \Lop\|\le N^{(\alpha_1-\alpha_r)/2}O_p(1/\sqrt{T})+\frac 1  {N^{(1+\alpha_r)/2}} {O_p(1)} $.
Squaring  gives the desired result.

The thrust of Proposition \ref{prop:prop6} is  that as far as consistent estimation of the factor space is concerned, the only loading strength that matters is that of the weakest, $\alpha_r$. Provided that $\alpha_r>0$, and  $N^{1-\alpha_r}/T\rightarrow 0$, the average error in estimating the factor space will vanish, albeit at a slower rate than in the strong loadings case.

\subsection{Distribution Theory}
We will use $\bm H_{NT,3}$ to obtain distributional results but we first need to establish its relation to $\bm H_{NT}$ when $\alpha$ is heterogeneous.

\begin{lemma} The following holds under Asumptions A',
\label{lem:H3=H0}
\begin{itemize}
\item[i.]
$ \bm H_{NT,3}-  \bm H_{NT}  = N^{\frac 1 2 \alpha_1 -\alpha_r}O_p(1)+\frac { N^{1+\frac 1 2 (\alpha_1-3\alpha_r)} }  {T} O_p(1)+  N^{\frac 1 2 (\alpha_1-3\alpha_r)} O_p(1)$.

\item[ii.] $ \bm H_{NT} \bm H_{NT,3}^{-1}=I_r  +N^{\frac 1 2 \alpha_1 -\alpha_r}O_p(1)+\frac { N^{1+\frac 1 2 (\alpha_1-3\alpha_r)} }  {T} O_p(1)+  N^{\frac 1 2 (\alpha_1-3\alpha_r)} O_p(1) $
\end{itemize}
\end{lemma}
The lemma  says $\bm H_{NT,3}-  \bm H_{NT}  = o_p(1)$ and $ \bm H_{NT} \bm H_{NT,3}^{-1}=I_r+o_p(1)$ if $\alpha_r>\alpha_1/2$ and $T$ is sufficiently large.  We will use this result to prove consistency of $\tilde C_{it}$.

To derive  the limiting distributions, we modify Assumptions B and C with the following:

\paragraph{Assumption B'.}  The following holds  for each $i$ and $t$ as  $N, T\rightarrow \infty$:
\begin{eqnarray*}
   \Bi \sum_{i=1}^N \Lambda^0 _i e_{it}\dconv \mathcal N(0,\bm \Gamma_t), \quad\quad
\frac{1}{\sqrt{T}}\sum_{t=1}^TF^0_te_{it}\dconv \mathcal N(0,\bm \Phi_i).
\end{eqnarray*}

\paragraph{Assumption C':}
(i) $   \alpha_r >\frac 1 2$, (ii)  $  \frac {\sqrt{T}} {N^{\alpha_r}} \rightarrow 0$, (iii) $   \frac  1{\sqrt{T}} \frac N {N^{\alpha_r}} \rightarrow 0$, and (iv) $
\frac 1 T \frac{N^{3/2}}{N^{\alpha_r}} \rightarrow 0$.

\bigskip
For the limiting distribution of $\tilde F_t$,  from the $t$-th row of (\ref{eq:FBFoB}), we have
\[ \B( \tF_t-\bH_{NT}' \Fo_t) = (N \B^{-2} \DNTr^2)^{-1} (\Bi \tF'\Fo \B/T)  \Bi \bm \Lambda^0 \e_t
+ (N \B^{-2} \DNTr^2)^{-1}
 \Bi \tF' \e \e_t/T. \]
 Now $(N \B^{-2} \DNTr^2)^{-1}\pconv \bm D_r^{-2}$, $(\Bi \tF'\Fo \B/T)\pconv \bm Q$. Furthermore,
 by Assumption B', $ \Bi \bm \Lambda^0 \e_t\dconv N(0,\bm \Gamma_t)$. The first  term is thus asymptotically normal. As shown in the Appendix,
 the second term  is
 \begin{equation}\label{eq:Feet}
\Bi \tF '\e \e_t/T=  \frac {N^{3/2}} {TN^{\alpha_r}}O_p(1) + N^{\frac 1 2 -\alpha_r} O_p(1)
\end{equation}
which is $o_p(1)$  under Assumption C'. So we have
 \[ \B( \tF_t-\bH_{NT}' \Fo_t)\dconv \mathcal N( 0, \bm D_{r}^{-2}\bm Q\bm \Gamma_t\bm Q'\bm D_{r}^{-2}). \]

 For  the distribution of  $\tilde\Lambda_i$,
\begin{equation} \label{eq:hat-loading}  \sqrt{T}(\tilde \Lambda_i -\bm H_{NT,3}^{-1} \Lambda_i^0) = \bm H_{NT}' \frac{ 1}{\sqrt{ T}} \sum_{t=1}^T F_t^0 e_{it} + \frac{\Bi(\tF\B-\Fo\bm H_{NT} \B)^\prime \e_i}{\sqrt{T}}.\end{equation}
Multiplying   by   $\bm H_{NT}^{\prime-1} $ on each side,
\[ \sqrt{T}\bm H_{NT}^{\prime-1} (\tilde \Lambda_i -\bm H_{NT,3}^{-1} \Lambda_i^0) =  \frac{ 1}{\sqrt{ T}} \sum_{t=1}^T F_t^0 e_{it} + \frac{\bm H_{NT}^{\prime-1} \Bi(\tF\B-\Fo\bm H_{NT} \B)^\prime \e_i}{\sqrt{T}}.\]
The first term is asymptotically normal by Assumption B'.  Using Lemma \ref{lem:limit}(ii) in appendix, the second  term is bounded by
\begin{equation*}  \|\frac{\bm H_{NT}^{\prime-1} \Bi(\tF\B-\Fo\bm H_{NT} \B)^\prime \e_i}{\sqrt{T}}\|\le  \frac {\sqrt{T}} {N^{\alpha_r}} O_p(1)
+N^{\frac{\alpha_1-2\alpha_r}2 }O_p(1)+ \frac{ N^{1-\alpha_r} } {\sqrt{T}} O_p(1) \end{equation*}
which is  $o_p(1)$ under Assumption C'.   Summarizing results we have

\begin{proposition} \label{prop:CLT2}
Suppose that  Assumptions A', B', and C' hold.  Then
\begin{itemize}
\item[i.] $ \bm B_N (\tilde F_t-\bm H_{NT}'F^0_t)\dconv \mathcal N(0, \bm D_{r}^{-2}\bm Q\bm \Gamma_t\bm Q'\bm D_{r}^{-2})$;
    \item[ii.] $\sqrt{T}\bm H_{NT}^{\prime-1} (\tilde \Lambda_i-\bm H_{NT,3}^{-1}\Lambda_i^0)\dconv \mathcal N(0,\bm \Phi_i  ).$
\label{prop:prop7}
\end{itemize}
\end{proposition}

 Part (i) of the Proposition says that  $\tilde F_{1t}$ associated with the strongest loading  will  converge to the normal distribution  at a faster rate of $\sqrt{\alpha_1}$ than the  $\tilde F_{rt}$ which only converges at rate $\sqrt{\alpha_r}$.




Assumption C' is needed for asymptotic normality but is stronger than is necessary for individual consistency of $\tilde \Lambda_i$ and $\tilde F_t$. All that is needed for  $\tilde \Lambda_i$ to be consistent is  $\alpha_r>0$. To see this, we can  divide (\ref{eq:hat-loading}) by $T^{1/2}$. The first term becomes
$\frac 1 {\sqrt{T}}\|\bH_{NT}\|O_p(1) = \sqrt{\frac {N^{\alpha_1}}{T N^{\alpha_r}} } O_p(1)=o_p(1)$. Now $\|\Bi\|=N^{-\alpha_r/2}$, and the second term is equal to $\Bi$ multiplied by the term
analyzed in Lemma \ref{lem:limit}(i) in the Appendix,
and this term  is $o_p(1)$ provided $\alpha_r>0$.
Similarly, we only  require $\alpha_r>1/3$ (together with $\frac {N^{3/2}} {TN^{3\alpha_r/2}}\rightarrow 0$)  for  $\tilde F_t$  to be consistent.
To see this,  we first multiply (\ref{eq:Feet}) by $\Bi$, which is $O(N^{-\alpha_r/2})$. Then
\[
\B^{-2}  \tF '\e \e_t/T= \frac {N^{3/2}} {TN^{3\alpha_r/2}}O_p(1) + N^{\frac {1  -3\alpha_r} 2} O_p(1). \]
Thus if $\alpha_r>1/3$ and $\frac {N^{3/2}} {TN^{3\alpha_r/2}}\rightarrow 0$,  the above is $o_p(1)$, which implies
$ \tF_t-\bH_{NT}' \Fo_t =o_p(1)$.   Both results are similar to the homogenous case, but with $\alpha$ replaced by $\alpha_r$.

 We next consider estimating the common component. Adding and subtracting terms,
\begin{eqnarray*}
\tilde C_{it}-C_{it}^0&=& \tilde\Lambda_i' \tilde F_t- \Lambda_i^{0'}F_t^0\\
&=&(\tilde \Lambda_i- \bm H_{NT,3}^{-1} \Lambda_0^i)'\tilde F_t  + \Lambda_i^{0'} \bm H_{NT,3}^{'^{-1}} (\tilde F_t- \bm H_{NT}' F_t^0)
+ (\Lambda_i^{0'} \bm H_{NT,3}^{'^{-1}}\bm H_{NT}' F_t^0 -\Lambda_i^{0'}F_t^0 )
\end{eqnarray*}
The first two terms are both  $o_p(1)$  by Proposition \ref{prop:CLT2}. The last term is $o_p(1)$ by part (ii) of
Lemma \ref{lem:H3=H0}.



We close this section with some remarks about our results in relation to those in the literature. The PC estimator imposes the normalization restrictions that
$\bm \Lambda'\bm \Lambda$ is diagonal, and $\bm F'\bm F/T=I_r$. If the true data generating process coincides with these  restrictions, that is,  $\Lop\Lo$ being diagonal and $\Fop\Fo/T=I_r$, then all the  rotation matrices introduced in this paper will be asymptotically an identity matrix as shown in \citet{baing-joe:13}.
Our analysis above has refrained from making the assumption that the true DGP coincides with the normalization restrictions so that we  can compare $\tF$ to $ \bm F^0\bm H_{NT}$, but not  of  $\tF_k$ to $\bm F_k$. The challenge  lies in being able to  define $\bm H_{NT}$ in a way  consistent with the data generating process so that $\frac{1}{T}\|\tF-\bm F^0\bm H_{NT}\|^2$ remains the metric for assessing estimation error.

Our approach contrasts with a growing body of work on this problem.
 \citet{uematsu-yamagata:inf, uematsu-yamagata:est}  consider regularized estimation and inference of sparsity induced weak factors.
  They assume in our notation that $ E[\bm H \bm F^0_t \bm F^{0'}_t\bm H']=\bm I_r$ and $\bm H^{-1\prime} \Lop  \Lo  \bm H^{-1}$ is a diagonal matrix  with   $\bm D_{jj}^2 N^{\alpha_j}$ in the $j$-th diagonal, which implicitly make unspecified restrictions about $\bm H$  and $\bm F^0$. Even as it is,
the average error bound for their estimator of $\bm F$ already restricts the relation between $\alpha_1$ and $\alpha_r$.\footnote{ They require that    $\alpha_1+\max(1, \tau)/2<3 \alpha_r/2+\tau/2$ where  $T=N^{\tau}$ for some $\tau>0$. As noted in their Remark 1,  the upper bound of $\alpha_1-\alpha_r$ of  1/4 is  attainable  when $\alpha_1=1$ with $\tau=(3/4,1]$, the  same as PC.
Their lower bound of $\alpha_r$ is attained  when $\alpha_1=\alpha_r$ and $\tau=2/3$,}  Their assumptions applied to PC estimation yields a strict lower bound of $\alpha_r> 1/2$,  stronger than the $\alpha_r>0$ result that we obtain. As the authors noted, without the implicit restrictions on $\bm H$ and $\bm F^0$ that are not innocuous,  additional assumptions on  $[\alpha_1,\alpha_r]$ would otherwise be required.


Most related to our result is  \citet{freyaldenhoven:22} whose goal is to   determine the number of (local) factors  whose  loadings are of varying strength. He assumes   $\alpha_k>1/2$,   $N/T\rightarrow c$,   $\bm F^{0'}\bm F^0/T$ is truly an identity matrix and  $\bm \Lambda^{0\prime}\bm \Lambda^0$ is truly diagonal with the implication that $\bm H$ is also an identity matrix. This simplifies the analysis as $\tF_{kt}$ estimates $\bm F_{kt}$, not just a rotation of it. With this additional identifying assumption, he finds that  $\alpha_k>1/2$  is required for consistency of $\tF_{kt} $ for $\bm F_{kt}^0$. For consistency, we obtain the result of  $\alpha_r>1/3$ without restricting $\bm H$ to be an identity matrix.


\section{Simulation Experiments}

To verify the asymptotic results, we conduct two simulation experiments with 5000 replications of data $X^0_{it}=\Lambda_i^{0'}F^0_t+\sigma_i e^0_{it}$ with $r=3$.  Let $\bm D^2$ and $\bm B$ are diagonal matrices,
$B_{jj}=N^{\alpha_j/2}$, $(j=1,2,3$).
Two data generating processes are considered.
\begin{itemize}
\item[i.]  DGP1: $\bm F^0=\sqrt{T}\bm U\bm D$ and $\bm \Lambda^0= \bm V\bm B$ where $\bm U$ and $\bm V$ are random orthonormal $T\times r$ and $N\times r$ matrices respectively. Hence $\bm F^{0'} \bm F^0/T=\bm D^2$  and  $\bm B^{-1} \bm \Lambda^{0'} \bm \Lambda^0 \bm B^{-1}$ is $\bm I_r$.
\item[ii.] DGP2,   $F_t^0\sim N(0,I_3)$,  $ \Lambda^0_i\sim N(0,I_r) \bm D \bm B/\sqrt{N}$. Hence $\bm F^{0'}\bm F^0/T\approx I_r$ and \newline $\bm B^{-1} \bm \Lambda^{0'}\bm \Lambda^0\bm B^{-1} \approx \bm D^2$ is diagonal.
\end{itemize}
   The common component $C_{it}^0=\Lambda_i^{0\prime} F_t^0$ has variance $\sigma^2_{Ci}$ calculated for each $i$ from the time series data.  The importance of the common component is $\bar R^2_C$, defined as the ratio of the mean of  the variance of the common component of each series to the mean of the variance of each series.
\begin{center}
\begin{tabular}{l|lll|lll} \hline
& \multicolumn{3}{c}{DGP1} & \multicolumn{3}{c}{DGP 2} \\ \hline
&  strong   & weak         & weak        & strong   & weak         & weak   \\
&            & homogenous & heterogenous &           & homogenous & heterogenous \\ \hline $D^2$ & \multicolumn{3}{c|}{$D^2=\diag(6\;  5\; 4)$} & \multicolumn{3}{c}{$D^2=\diag(3\; 2\; 1)$}\\
$ \alpha_1,\alpha_2,\alpha_3$ &  (1,1,1)  &      $(.25,.25,.25) $  &       $(1,1/3,1/6 )$&
  (1,1,1)  &      $(.4,.4,.4) $  &       $(1,2/3,1/3 )$\\
$\bar R^2_C $ & 0.541 & 0.082 & 0.327 &  0.545 & 0.109 & 0.50 \\
\hline
\end{tabular}
\end{center}
By design of $\bm B$, $F_1^0$ contributes more  than $F_2^0$ and $F_3^0$ to the variations in the data.
In the strong factor case, the parameterizations yield a common component that accounts for a bit over half of the total variations in the data in both DGPs and reduces to less than 10\% when all loadings are equally weak.  In the heterogeneous case, strong and weak loadings co-exist and the common component explains about one-third of the total variations in DGP1 and half in DGP2.

For each factor $j=1,\ldots, r$, we regress the $j$-th column of $\tilde F$ on $F^0$. The coefficients estimate the $j$-th column of $\bm H_{NT,3}$, so the $R^2$ from the regression is an assessment of fit. Likewise, the $j$-th column of $\tilde \Lambda$ is regressed on $\Lambda^0$ and the coefficients estimate the $j$-th row of $\bm  H^{-1}_{NT,4}$. The residuals from these regressions are then the (non-normalized) estimation error. Tables \ref{tbl:table1} and  \ref{tbl:table2}  report both the $R^2$ for each $j$, as well as $M(\tilde{\bm F})$=trace(top)/trace(bottom), a multivariate measure of fit between $\bm F^0 $ and $\tF$, where top=
$\bm F^{0'} \tilde{\bm F} (\tilde {\bm F}^{'}\tilde{\bm F})^{-1} \tilde {\bm F'}\bm F^0$ and bottom = $\bm F^{0'}\bm F^0$. The statistic $M(\tL)$ is similarly defined. The last statistic is the average of correlation between $\tilde { C}_i$ and  $ C^0_i$. The distributions of the estimation error for $\tilde F_t$  are shown in Figure \ref{fig:fig1} for $t=100$. Figure  \ref{fig:fig2} shows the distributions of the estimation error for $\tilde \Lambda_i$ at $i=50$. In both cases, $T=500$ and $N=100$. Both distributions appear symmetric.

We consider eight configurations of $N$ and $T$. In the strong factor case, all statistics indicate that the factors are precisely estimated.  Under Assumption A4, $N^{1-\alpha_r}/T$ must tend to zero. Hence for given $N$, a larger $T$ will give faster convergence of the estimates to a rotation of the true values. The results bear this out. In the homogeneous $\alpha$ case reported in the middle panel,  the estimation errors are similar across factors. This is different for the weak-heterogeneous case in the bottom panel as estimates of $\bm F_1$ and $\bm \Lambda_1$ are more precise than for $\bm F_3$ and $\bm \Lambda_3$, as suggested by theory.  The common component remains precisely estimated with errors closer to the stronger loadings case than the weak-homogeneous case because the dominant factor is strong.

So far, DGP1 assumes that the factors are orthogonal and DGP2 assumes that the factors are asymptotically orthogonal. Since our theory does not require the assumption that $\bm F^{0'}\bm F^0$ is a diagonal matrix, we also modify DGP1 so that $U$ and $V$ are no longer orthogonal. This is achieved by multiplying these orthogonal vectors into two $r\times r$ matrices with random elements below the diagonal. The results, shown in Table \ref{tbl:table3}, are similar to Table \ref{tbl:table1}.

\section{Conclusion}
A sizable literature has emerged since \citet{onatski-joe:12} shows that the PC estimates are inconsistent when the loadings are extremely weak in the sense that  $\frac{\bm\Lambda'\bm \Lambda}{N^\alpha} \rightarrow \bm \Sigma_\Lambda$ with $\alpha=0$.  This paper establishes conditions under which  the PC estimates  are consistent and asymptotically normal in more moderate cases when $1\ge \alpha > 0$.
The main conclusion is unchanged in  the heterogeneous $\alpha$ case, where the one $\alpha$ that determines consistency is that of the weakest loading. The takeaway is that asymptotic normality requires $\alpha_r>1/2$, stronger than is needed for the consistency results.

Allowing  $\alpha$ to take on a range of values  naturally raises questions about the different criteria available to determine the number of factors.  If we want to estimate the number of factors with $\alpha>0$, the  criteria  in \citet{baing-ecta:02} remain useful. While the criteria of \citet{onatski-restat} and  \citet{ahn-horenstein:13}  try to better separate the bounded from the diverging eigenvalues of $\bm X\bm X'$,  \citet{baing-joe:19}  seek to isolate the factors with a tolerated level of explanatory power by singular value  thresholding.  This  criteria will  return an estimate of the `minimum rank', a concept of long standing interest  in classical factor analysis, see, e.g., \citet{tenberge-kiers}.  For documenting the number of factors with different strength, methods of  \citet{bailey-kapetanios-pesaran:16} and  \citet{uematsu-yamagata:est} are available. The criteria in \citet{freyaldenhoven:22} determine the number factors with $\alpha_r>1/2$. But as seen above, the $\alpha$ required for consistent estimation of the factor space can be smaller than the one needed for asymptotic normality. Ultimately,
 the desired number of factors    depends on the objective of the exercise and the assumptions that the researcher find defensible.   It seems difficult to avoid taking a stand on  what is meant by  weak in practice.

\newpage
\section*{Appendix}
\setcounter{lemma}{0}
\setcounter{equation}{0}
\setcounter{section}{0}
\renewcommand{\thelemma}{A.\arabic{lemma}}
\renewcommand{\theequation}{A\arabic{equation}}
\renewcommand{\thesection}{A\arabic{section}}
The following inequalities are used to simplify the proofs in earlier work.
\begin{lemma}
\label{lem:usefulprelim}
The following holds under Assumption A:
\begin{subequations}
\begin{eqnarray}
 \| \bm e '\bm F^0\|^2  & =&   \tr(\Fop \bm e  \bm e ' \Fo ) =O_p(NT)  \label{eq:6a} \\
 \|\bm e \bm \Lambda^0 \|^2  & = &  \tr( \bm \Lop \bm e ' \bm e  \Lo )  =O_p(N^\alpha T) \label{eq:6b} \\
 \|\bm e  \bm e' \bm F^0\|  &\le&   \rho_{\max} (\bm e \bm e ') \|\bm F^0\| =O_p(\max\{N,T\} \sqrt{T}) \label{eq:6c} \\
 \| \bm e ' \bm e  \Lo\| &\le&  \rho_{\max}(\bm e \bm e^\prime)\|\bm \Lambda^0\|=  O_p(\max\{N, T\} \sqrt{N^\alpha}) \label{eq:6d}
\end{eqnarray}
\end{subequations}
\end{lemma}
The matrix norm $\|\cdot \|$  here is the  Frobenius norm. Since the matrices are rank $r$ (fixed), spectral norms would give the same bounds.

\paragraph{Proof of Lemma \ref{lem:usefulprelim}:} (\ref{eq:6a}) and (\ref{eq:6b}) follow from  (\ref{eq:FeeF}) and
(\ref{eq:LeeL}), respectively. For
(\ref{eq:6c}),
\[ \|\bm e  \bm e' \bm F^0\|  \le  \|\bm e \bm e '\|_{sp} \|\bm F^0\|
=   \rho_{\max} (\bm e \bm e ') \|\bm F^0\|=O_p(\max\{N,T\} \sqrt{T}) \]
here we used $\|\bm A \bm B\|\le \|\bm A\|_{sp} \|\bm B\|$.
The argument for (\ref{eq:6d}) is the same.

\section*{Proof of Results in Sections 3}

\paragraph{Proof of Lemma  \ref{lem:lemmaH} part (i)}
We start with (\ref{eq:identity3}). We derive the order of magnitude for each of the  three terms in braces.
\begin{eqnarray*}
 \frac{\bm \Lop \bm e^{\prime} \tF}{N^\alpha T} &=& \frac{\bm \Lop \bm e^{\prime} \bm \Fo}{N^\alpha T}
 + \frac{\bm \Lop \bm e^{\prime} (\tF-\bm \Fo H_{NT,0}) }{N^\alpha T},\\
 \|\frac{\bm \Lop \bm e^{\prime} \tF}{N^\alpha T}\| & \le & \|\frac{\bm \Lop \bm e^{\prime} \bm \Fo}{N^\alpha T}\|
 + \|\frac{\bm \Lop \bm e^{\prime}}{N^\alpha \sqrt{T} }\|  \frac{ \| (\tF-\bm \Fo H_{NT,0})\| }{ \sqrt{T} }\\
 &=& O_p(\frac 1 {\sqrt{N^\alpha T}})+  O_p(\frac 1 {\sqrt{N^\alpha}}) \Big[O_p(\frac 1 {\sqrt{N^\alpha}})+\frac 1 {T}  \frac N {N^\alpha} O_p(1)\Big]\\
  &=& O_p(\frac 1 {\sqrt{N^\alpha T}})+  O_p(\frac 1 {N^\alpha}) + \frac 1 { T}  \frac N {N^{3\alpha/2}} O_p(1),
 \end{eqnarray*}
 where we have used Assumption A3(v), (\ref{eq:6b}), and Proposition \ref{prop:prop1}. Next (\ref{eq:FeeF-hat}) implies
\begin{eqnarray*}\|\frac{\bm \tF'\bm e \bm e^{\prime}\tF}{N^\alpha T^2}\|
& = & (\frac N {N^{\alpha}}) \frac 1 T  O_p(1) +\frac 1 {N^{\alpha}} O_p(1).
\end{eqnarray*}
Collecting the non-dominated terms, we obtain
\begin{eqnarray*}
&& \Big(\frac{\bm \tF' \bm F^0}{T}\Big) \bm H_{NT,0}  =  \bm I_r +O_p(\frac 1 {\sqrt{N^\alpha T}})+  O_p(\frac 1 {N^\alpha})
 +(\frac N {N^{\alpha}}) \frac 1 T  O_p(1).
\end{eqnarray*}
Equivalently,
 \begin{eqnarray*}
 &&  \bm H_{NT,0}  =  \Big(\frac{\bm \tF' \bm F^0}{T}\Big)^{-1}+O_p(\frac 1 {\sqrt{N^\alpha T}})+  O_p(\frac 1 {N^\alpha})
 +(\frac N {N^{\alpha}}) \frac 1 T  O_p(1).
\end{eqnarray*}
$\Box$

Part (i) makes clear that the approximations for $\bm H_{NT,0}$ depends on $\alpha$.

\noindent {\bf Proof of Lemma \ref{lem:lemmaH} part (ii):}
  We already proved the case of $\ell=3$. It remains to show  asymptotic equivalence of the three remaining matrices to $\bm H_{NT,0}$.

We begin with $\ell=1$ where $\bm   H_{NT,1}=(\Lp\bm \Lambda^0/N^\a)(\tL\pr \bm \Lambda^0/N^\a)^{-1}$. From $\bm \tLp=\tFp \bm X/T=\tFp (\bm F^0 \bm \Lop +\bm e)/T$, we have
\begin{eqnarray} \label{eq:tLpLo} \frac{\bm \tLp \bm \Lo} {N^\a} =\frac{\tFp \bm F^0}{T} \frac { \bm \Lop \bm \Lo}{N^\a}  +\frac { \tFp \bm e \Lo} {TN^\a}
=  \frac{\tFp \bm F^0}{T} \frac { \bm \Lop \bm \Lo}{N^\a} + \frac 1 {\sqrt{T N^\a}} O_p(1).
\end{eqnarray}
Thus
$ \bigg ( \frac{\bm \tLp \bm \Lo} {N^\a} \bigg)^{-1} = \bigg( \frac{\bm \Lop \bm \Lo}{N^\a}\bigg)^{-1} \bigg(\frac{\tFp \bm F^0}{T} \bigg)^{-1}+\frac 1 {\sqrt{T N^\a}} O_p(1) $, implying
\[ \bm H_{NT,1}=\Big(\frac{\tFp \bm F^0}{T} \Big)^{-1}+\frac 1 {\sqrt{T N^\a}} O_p(1). \]
 For $\ell=4$, taking the transpose of (\ref{eq:tLpLo}) gives
\begin{eqnarray*} \bm H_{NT,4}&=& \frac{\bm \tLp \bm \Lo} {N^\a} \frac {N^\a} N \bm D_{NT,r}^{-2}=
\bigg(\frac{\bm \Lambda^{0^\prime}\bm \Lambda^0}{N^\a}\bigg)\bigg(\frac{\bm F^{0^\prime}\tF}{T}\bigg) \frac {N^\a} N \bm D_{NT,r}^{-2}+
\frac 1 {\sqrt{T N^\a}} O_p(1) \\ &=&\bm H_{NT,0}+   \frac 1 {\sqrt{T N^\a}} O_p(1)  .
\end{eqnarray*}
Finally, for $\ell=2$, we use the expression for $\tF$ in (\ref{eq:factorspace}) to obtain
\[ \bm H_{NT,2} - \bm H_{NT,0} \equiv   \Big( \frac{\Fop \Fo} T\Big)^{-1}   \Big(\frac{\Fop \Fo\Lop\bm e^{\prime} \tF}{N^\alpha T^2} +\frac{\Fop \bm e \Lo \Fop\tF}{N^\alpha T^2}+ \frac{\Fop \bm e \bm e^{\prime}\tF}{N^\alpha T^2} \Big)  \Big( \frac N {N^\alpha} \bm  D_{NT,r}^{2}\Big)^{-1} \]
The right hand side is bounded by $ O_p(\frac 1 {\sqrt{N^\alpha T}})+  O_p(\frac 1 {N^\alpha})
 +(\frac N {N^{\alpha}}) \frac 1 T  O_p(1)$.   $\Box$

\noindent {\bf Proof of Lemma \ref{lem:lemmaH} part (iii):}
 Part (iii) follows from  parts (i) and (ii) and  $\bm H_{NT,0} \pconv \bm Q^{-1}$. $\Box$

\paragraph{Proof of Proposition \ref{prop:prop2}.} As argued in the text, it remains to analyze the term
$\frac{ \|(\tF- \Fo \bm H_{NT,0})' \bm e\| } {T\sqrt{N}} $. Notice
\begin{eqnarray} \label{eq:eFF0/T} \frac {\bm e'( \tF -\Fo \bm H_{NT,0})} T  &=&\Big(\frac{\bm e' \Fo\Lop\bm e^{\prime} \tF}{N^\alpha T^2} +\frac{\bm e' \bm e \Lo \Fop\tF}{N^\alpha T^2}+ \frac{\bm e' \bm e \bm e^{\prime}\tF}{N^\alpha T^2} \Big)  \Big( \frac N {N^\alpha} \bm  D_{NT,r}^{2}\Big)^{-1}
\end{eqnarray}
The right hand side $\tF$ can be replaced by $\Fo$ without affecting the result. In fact,
\[ \frac{\bm e' \Fo\Lop\bm e^{\prime} \tF}{N^\alpha T^2} =\frac{\bm e' \Fo\Lop\bm e^{\prime} \Fo}{N^\alpha T^2}
+o_p(1) \frac {\bm e'( \tF -\Fo \bm H_{NT,0})} T \]
because $\|\frac{\bm e' \Fo\Lop} {\Na T}\|=o_p(1)$. The second term above can be combined with the right hand side of (\ref{eq:eFF0/T}).
The same argument is applicable to $\frac{\bm e' \bm e \bm e^{\prime}\tF}{N^\alpha T^2}$  on the right hand side of  (\ref{eq:eFF0/T}). Replacing $\tF$ by $\Fo$, we have
\begin{eqnarray*} \|\frac{\bm e' \Fo\Lop\bm e^{\prime} \Fo}{N^\alpha T^2}\| &\le& \frac 1 {N^\alpha T^2} \|\e' \Fo\|\cdot \|\Lop \bm e'\Fo\| =\frac 1 T \sqrt{\frac N {N^{\alpha}}}  O_p(1) \\
 \|\frac{\bm e' \bm e \Lo }{N^\alpha T}\|&=& \frac { \max\{N,T\} \sqrt{N^{\alpha}} } {N^\alpha T} O_p(1)
\le  \frac N {\sqrt{N^\alpha}} \frac 1 T O_p(1) + \frac 1 {\sqrt{N^{\alpha}}} O_p(1) \\
 \|\frac{\bm e' \bm e \bm e^{\prime}\bm \Fo}{N^\alpha T^2} \| &\le& \frac 1 {N^\alpha T^2} O_p(\max\{N,T\}) \|\e'\Fo\|
 =\frac 1 {N^\alpha T^2} O_p(\max\{N,T\})  O_p(\sqrt{NT})\\ &=&\frac{N^{3/2-\alpha}}{T^{3/2}}O_p(1)
+  \sqrt{\frac N T} \frac 1 {N^\alpha} O_p(1).
\end{eqnarray*}
Collecting the non-dominating terms, we obtain
\begin{equation} \label{eq:FtFH0e/T} \frac{ \|(\tF- \Fo \bm H_{NT,0})' \bm e\| } {T} \le  \frac N {\sqrt{N^{\alpha}}} \frac 1 T  O_p(1) +
 \frac 1 {\sqrt{N^{\alpha}}} O_p(1)
 +\frac{N^{3/2-\alpha}}{T^{3/2}}O_p(1)
+  \sqrt{\frac N T} \frac 1 {N^\alpha} O_p(1)
\end{equation}
So
\[ \frac{ \|(\tF- \Fo \bm H_{NT,0})' \bm e\| } {T\sqrt{N}}  \le   \frac 1 T \sqrt{\frac N {N^{\alpha}}}  O_p(1) +
\frac 1 {\sqrt{N^{1+\alpha}}}
+\frac 1 { T^{3/2}}  (\frac N {N^{\alpha}}) O_p(1) + \frac 1 {N^\alpha  \sqrt{T}} O_p(1).\]

$\Box$

The following lemma is used to prove  results in Section 4, i.e., the limiting distributions.

\begin{lemma}
\label{lem:lemma6}
Suppose that Assumption A holds.  We have, for $\ell=0,1,2,3,4$,

\begin{itemize}
\item[i] $  \frac 1 T \Fp (\tF- \Fop \bm H_{NT,\ell})   = O_p(\frac 1 {\sqrt{N^\alpha T}})+O_p(\frac N {N^\alpha T})+ O_p(\frac 1 \Na)+
 \frac 1 {N^{\alpha}} (\frac N {T\Na})^{1/2} O_p(1)  $
\item[ii]
$   \frac 1 \Na  \Lp(\tL-\Lo\bm H_{NT,\ell}^{\prime -1})   = O_p(\frac 1 {\sqrt{N^\alpha T}})+ O_p(\frac 1 \Na)+ \frac 1 {\sqrt{N^\alpha}} O_p\left(\frac N {T\Na}  \right)$
\item[iii]  $   \frac 1 T e_i' (\tF- \Fo \bm H_{NT,\ell})   =O_p(\frac 1 {N^\alpha} ) + O_p(\frac 1 {\sqrt{T\Na}})+O_p(\frac N {N^{\alpha}T} )$,
for each $i$,
\item[iv]  $  \frac 1 N e_t'(\tL-\bm \Lo \bm H_{NT,\ell}^{\prime -1})   = \frac 1 T \sqrt{\frac N {N^{\alpha}}}  O_p(1) +
\frac 1 {\sqrt{N^{1+\alpha}}} O_p(1)
+\frac 1 { T^{3/2}}  (\frac N {N^{\alpha}}) O_p(1) + \frac 1 {N^\alpha  \sqrt{T}} O_p(1)$,
  for each  $t$.
\end{itemize}

\end{lemma}
The  same bounds in parts (i) and (ii) hold for $\frac 1 T \tF' (\tF- \Fop \bm H_{NT,\ell})$
and  $\frac 1 \Na  \tLp(\tL-\Lo\bm H_{NT,\ell}^{\prime -1})$. This follows from, for example,
$\frac 1 T \tF' (\tF- \Fop \bm H_{NT,\ell})= \bm H_{NT}'\frac 1 T \Fp (\tF- \Fop \bm H_{NT,\ell})+\frac 1 T \|\tF- \Fop \bm H_{NT,\ell}\|^2$
and Proposition \ref{prop:prop1}.
In the strong factor case, all four quantities are $O_p(\delta_{NT}^{-2})$.
The rates now depend on $\alpha$ and affect the convergence rates.

\paragraph{Proof of Lemma \ref{lem:lemma6}.} Consider (i).
\begin{eqnarray*}
 \frac 1 T \Fop( \tF -\Fo \bm H_{NT,0})
   &=&\Big[\Big(\frac{\Fop\Fo} T \Big) \Big(\frac{\Lop\bm e^{\prime} \tF}{N^\alpha T}\Big) +\Big(\frac{\Fop\bm e \Lo}{N^\alpha  T}\Big) \frac{ \Fop\tF}{ T}+ \frac{\Fop \bm e \bm e^{\prime}\tF}{N^\alpha T^2}  \Big]  O_p(1)
   \label{eq:temp}
\end{eqnarray*}

Consider the first term in the bracket, and write $\boldsymbol H$ for $\boldsymbol H_{NT,0}$ (in fact result holds for all $\boldsymbol H_{NT,\ell}$
because of Lemma \ref{lem:lemmaH})
\[ \frac{\Lop\bm e^{\prime} \tF}{N^\alpha T}= \frac{\Lop\bm e^{\prime} \Fo \bm H }{N^\alpha T} +\frac{\Lop\bm e^{\prime}( \tF- \Fo \bm H)}{N^\alpha
 T}
\]
 the first term is $O_p(\frac 1 {\sqrt{N^\alpha T}})$ by Assumption A3(v). The second term is bounded by
\begin{eqnarray}
 \label{eq:Le(F-Fhat)}
& & \| \frac{\Lop\bm e^{\prime} ( \tF- \Fo \bm H)}{N^\alpha T} \|   \le  \|\frac 1 {\sqrt{N^\alpha}} (\frac{ \| \bm \Lambda^{0\prime}\bm e^{\prime}\| }{ \sqrt{\Na T} }) \|  \| \frac{( \tF- \Fo \bm H)}{\sqrt{T} }\|   \nonumber \\
& & =  \frac 1 {\sqrt{N^\alpha}} \Big[ O_p(\frac 1 {\sqrt{N^\alpha}} )+ O_p(\frac N{TN^\alpha})\Big]  =O_p\left( \frac 1 \Na\right) + \frac 1 {\sqrt{N^\alpha}} O_p\left(\frac N {T\Na}  \right)
\end{eqnarray}
Here we used Proposition \ref{prop:prop1}.



 The second term in the bracket $\frac{\Fop\bm e \Lo}{N^\alpha  T}=O_p(\frac 1 {\sqrt{\Na T}})$ by Assumption A3(v). Next
\begin{eqnarray} \label{eq:F*(F-Fhat)}
\|\frac{\Fop \bm e \bm e^{\prime}\tF}{N^\alpha T^2} \| & \le & \|\frac{\Fop \bm e \bm e^{\prime}\Fo}{N^\alpha T^2}\|  +\|\frac{\Fop \bm e\|}{\Na T} \| \frac{\bm e^{\prime}(\tF-\Fo H)}{ T}\| \nonumber \\
& = & \frac 1 T  O_p(\frac N \Na )+ \frac 1 \Na O_p(\sqrt{\frac N T}) \| \frac{\bm e^{\prime}(\tF-\Fo H)}{ T}\| \nonumber\\
& = & \frac 1 T  O_p(\frac N \Na )+\left(\frac N {T\Na}\right)^{3/2}O_p(1)+\frac 1 {N^\alpha} O_p( \sqrt{\frac N {N^\alpha T}})+\left(\frac N {T\Na}\right)^2 O_p(1) +\frac 1 {\Na} O_p\left(\frac N {T\Na}\right)  \nonumber \\
 & = & \frac 1 T  O_p(\frac N \Na )+\frac 1 {N^\alpha} O_p( \sqrt{\frac N {N^\alpha T}})
\end{eqnarray}
Here the second equality uses (\ref{eq:FtFH0e/T}). The last equality keeps the dominant terms because $N/(T\Na)\rightarrow 0$. Finally,
collecting the non-dominated terms gives (i).
 None of these terms can be dominated by others, all depending on the relative magnitude of $N$ and $T$, and
when $\alpha=1$, they simplify to $O_p(\delta_{NT}^{-2})$.

Consider (ii).
\[ \frac 1 \Na (\tL-\Lo\bm H_{NT,3}^{\prime -1})' \Lo = \bm H_{3,NT}' \Fop \bm e \Lo /{(\Na T)} + (\tF- \Fo \bm H_{NT,3})' \bm e \Lo /{(\Na T)} \]
the first term is $O_p(\frac 1 {\sqrt{\Na T} })$ by Assumption A3(v), the second term is analyzed in (\ref{eq:Le(F-Fhat)}). This proves (ii).

Consider (iii)

\begin{eqnarray*} \frac { e_i'( \tF -\Fo \bm H_{NT,0})} T  &=&\Big(\frac{ e_i' \Fo\Lop\bm e^{\prime} \tF}{N^\alpha T^2} +\frac{e_i' \bm e \Lo \Fop\tF}{N^\alpha T^2}+ \frac{ e_i' \bm e \bm e^{\prime}\tF}{N^\alpha T^2} \Big)  \Big( \frac N {N^\alpha} \bm  D_{NT,r}^{2}\Big)^{-1}
\end{eqnarray*}
The first term can be ignored. The second term is determined by
\[ \frac{e_i' \bm e \Lo}{N^\alpha T} =O_p(\frac 1 {N^\alpha})+ O_p(\frac 1 {\sqrt{T N^\alpha}}) \]
by Assumption A3(ii).
The third term is $O_p(\frac 1 {N^\alpha})+ \frac N {N^\alpha T})$. Collecting terms gives (iii).

Consider (iv).

\[ \frac 1  N (\tL-\Lo\bm H_{NT,3}^{\prime -1})'e_t = \bm H_{3,NT}' \Fop \bm e e_t /{(N T)} + (\tF- \Fo \bm H_{NT,3})' \bm e e_t /{(N T)} \]
First term is $O_p(\delta_{NT}^{-2})$ by Assumption A3. The second term is bounded by
\[ \frac{\|(\tF- \Fo \bm H_{NT,3})' \bm e\|}{T\sqrt{N}}   \frac{\|e_t\|} {\sqrt{N}}  =\frac 1 T \sqrt{\frac N {N^{\alpha}}}  O_p(1) +
\frac 1 {\sqrt{N^{1+\alpha}}}O_p(1)
+\frac 1 { T^{3/2}}  (\frac N {N^{\alpha}}) O_p(1) + \frac 1 {N^\alpha  \sqrt{T}} O_p(1)\]
This proves (iv).  $\Box$

\section*{Proof of Results in Section 5}
We shall use $\bar \bH$ and $\bar \bH_{NT}$ interchangeably.

\noindent {\bf Proof of  (\ref{FBFBH})}.  From (\ref{eq:FBFoB})
\begin{eqnarray*} \label{eq:FBFoB1}\tF \B-\Fo \B \bar \bH & = & \frac{\bm e\bm \Lambda^0  \bm  F^{0^\prime}\tF \Bi}{T}(N \B^{-2} \DNTr^2)^{-1}
+ \frac{\bm e \bm e^{\prime}\tF \Bi}{ T} (N \B^{-2} \DNTr^2)^{-1}  \\
 &=& a + b, \quad \text{say} \nonumber
 \end{eqnarray*}
where
\begin{eqnarray*} \|a\| &\le& \|\bm e\bm \Lambda^0\Bi\| \| \frac {\B F^{0^\prime}\tF \Bi}{T}\|\|(N \B^{-2} \DNTr^2)^{-1} =
\|\bm e\bm \Lambda^0\Bi\| O_p(1) =O_p(\sqrt{T}) \\
 \|b\| &\le& \frac {\max(N,T)} T \|\tF\| \|\Bi\| O_p(1)= \frac {\max(N,T)} {\sqrt{T}}  N^{-\alpha_r/2} O_p(1)
\end{eqnarray*}
$\Box$
\bigskip

\noindent {\bf Proof of (\ref{eq:LHL})}. From (\ref{eq:FBFoB}),
$\tF \B-\Fo \B \bar \bH = \frac{\bm e\bm \Lambda^0  \bm  F^{0^\prime}\tF \Bi}{T}O_p(1)
+ \frac{\bm e \bm e^{\prime}\tF \Bi}{ T} O_p(1) $. Hence
\begin{align*}\bm e'(\tF \B-\Fo \B \bar \bH) & =  \frac{\bm e' \bm e\bm \Lambda^0  \bm  F^{0^\prime}\tF \Bi}{T}O_p(1)
+ \frac{\bm e' \bm e \bm e^{\prime}\tF \Bi}{ T} O_p(1)
  =  c +d, \text{ say}
\end{align*}
where
\begin{eqnarray*}
 \|c\|&\le& \|\be'\be \|_{sp}\|\Lo\Bi\|\|(\B \Fop\tF \Bi/T)\|
=\max\{N,T\} O_p(1) \\
 \|d\|&\le& \|\e\|_{sp}^3 \|\tF\|\|\Bi\|/T=\max\{N^{3/2},T^{3/2}\} \sqrt{T} N^{-\alpha_r/2}/T O_p(1) .
\end{eqnarray*}
Thus
\begin{equation}
\bm \|e'(\tF \B-\Fo \B \bar \bH)\|  =\max\{N,T\} O_p(1)+\max\{N^{3/2},T^{3/2}\}  N^{-\alpha_r/2}/\sqrt{T}O_p(1)  \label{eq:eFBFBH}
\end{equation}
and
\begin{eqnarray*}
\frac{ \|\Bi\| \|(\tF \B- \Fo \bm H_{NT} \B)' \bm e\| } {T\sqrt{N} } & \le &
\frac{\max\{N,T\}} T \frac {O_p(1)} {N^{(1+\alpha_r)/2}}+\frac {\max\{N^{3/2},T^{3/2}\}} {T^{3/2}} \frac {O_p(1)} {N^{\alpha_r+\frac 1 2}}   \\
& =&\frac 1  {N^{(1+\alpha_r)/2}} {O_p(1)} + o_p(\frac 1 {\sqrt{T}}).
\end{eqnarray*}
Note $N/(TN^{(1+\alpha)/2})=o_p(1/\sqrt{T})$ and $N/(T^{3/2} N^{\alpha_r})=o_p(1/\sqrt{T})$ under
$N/(TN^{\alpha_r})\rightarrow 0$. $\Box$

The following is needed to prove Proposition \ref{prop:prop7}.

\begin{lemma} \label{lem:limit} Under Assumption A' and B'
\begin{itemize}
\item[(i)] $  \frac 1 T (\tF \B-\Fo \B \bar \bH)'\bm e_i  =O_p(N^{-\alpha_r/2})+O_p(\sqrt{\frac {N^{\alpha_1}}{T N^{\alpha_r}} })+ N^{-\alpha_r/2}\frac N T O_p(1)$;
\item[(ii)]
$ \|\frac{\bm H_{NT}^{\prime-1} \Bi(\tF\B-\Fo\bm H_{NT} \B)^\prime \e_i}{\sqrt{T}}\|\le \sqrt{T} N^{-\alpha_r} O_p(1)
+N^{(\alpha_1-2\alpha_r)/2}O_p(1)+ \frac{ N^{1-\alpha_r} } {\sqrt{T}} O_p(1).$
\end{itemize}
\end{lemma}
Proof of (i):
\begin{align*} \frac 1 T \bm e_i'(\tF \B-\Fo \B \bar \bH) & =  \frac{\bm e_i' \bm e\bm \Lambda^0  \bm  F^{0^\prime}\tF \Bi}{T^2}O_p(1)
+ \frac{\bm e_i' \bm e \bm e^{\prime}\tF \Bi}{ T^2} O_p(1) \\
 & =  p_i +q_i, \text{ say}
\end{align*}
\[ \|p_i\|\le \frac 1 T  \|\e_i'\e \Lo \Bi\|\frac {\|\B \Fop \tF \Bi\| }{T}O_p(1) = O_p(N^{-\alpha_r/2})+O_p(\sqrt{\frac {N^{\alpha_1}}{T N^{\alpha_r}} }) \]
where we used Assumption A3'(iv).
\[ \|q_i\| \le \frac 1 {T^2} \|\e_i\| \max\{N,T\} \|\tF\| \|\Bi\| O_p(1) =N^{-\alpha_r/2} \max\{N/T,1\} O_p(1) \]
\[ \le N^{-\alpha_r/2}O_p(1)+ N^{-\alpha_r/2}\frac N T O_p(1) \]
since both $\|\e_i\|$ and $\|\tF\|$ are $O_p(T^{1/2})$.
This proves part (i).

\noindent Proof of (ii). From $\bH_{NT}=\B \bar \bH_{NT} \Bi$, we have $\bm H_{NT}^{\prime-1}\Bi =\Bi \bar  \bH_{NT}^{\prime-1}$. Thus $|\bm H_{NT}^{\prime-1}\Bi\| =\|\Bi \bar  \bH_{NT}^{\prime-1}\|
\le \|\Bi\| O_p(1) =O_p(N^{-\alpha_r/2})$.
Part (ii) is obtained by  multiplying  the bound in part (i) by $\sqrt{T} N^{-\alpha_r/2}$. This proves the lemma.
$\Box$

\bigskip
\noindent {\bf  Proof of (\ref{eq:Feet})}.  Adding and subtracting terms,
\[ \Bi \tF '\e \e_t/T=\Bi (\tF-\Fo  \bH_{NT})'\e \e_t /T +\Bi \bH_{NT}'\Fop \e \e_t/T \]
The second term is bounded by, using Assumption A3', $\|\Fop\e\e_t/T\|\le  N O_p(\delta_{NT}^{-2})$,
\begin{align*} \|\Bi \bH_{NT}'\Fop \e \e_t/T\|& \le \|\Bi\bH_{NT}'\| \|\Fop\e\e_t/T\|=\|\Bi \bH_{NT}'\|N O_p(\delta_{NT}^{-2})
\end{align*}
From $\Bi \bH_{NT}'= \B^{-2} \bar \bH' \B$, we have $\|\Bi \bH_{NT}'\|= \|\B^{-2}\| \|\bar \bH'\|\| \B\|= O_p( N^{(\alpha_1-2\alpha_2)/2})$.
It follows that
\begin{align*} \|\Bi \bH_{NT}'\Fop \e \e_t/T\|&
 =O_p(N^{(\alpha_1-2\alpha_r)/2}) N O_p(\delta_{NT}^{-2})=O_p(N^{(\alpha_1-2\alpha_r)/2}) + O_p(\frac N T) N^{^{(\alpha_1-2\alpha_r)/2}}.
\end{align*}
Next
\[ \Bi (\tF-\Fo \bH_{NT})'\e \e_t /T =\B^{-2} (\tF\B-\Fo\bH_{NT} \B)'\e\e_t/T  \]
Note that
$\bH_{NT}\B=\B \bar \bH$. By (\ref{eq:eFBFBH}) and since $\|e_t\|=O_p(N^{1/2})$,
\begin{align*} & \| \B^{-2} (\tF\B-\Fo \B \bar \bH)'\e\e_t/T\|  \le N^{-\alpha_r} \|(\tF\B-\Fo \B \bar \bH)'\e\|\|e_t\|/T  \\
& =N^{-\alpha_r} \max\{N,T\} N^{1/2}/T O_p(1)+N^{-\alpha_r} \max\{N^{3/2},T^{3/2}\}  N^{-\alpha_r/2}N^{1/2}/T^{3/2} O_p(1)\\
& = \frac {N^{3/2}} {TN^{\alpha_r}}O_p(1) + N^{\frac 1 2 -\alpha_r} O_p(1)
+\Big(\frac N {T N^{\alpha_r}}\Big)^{1/2}\frac {N^{3/2}}{T N^{\alpha_r}} O_p(1) +N^{\frac 1 2 -\frac 3 2 \alpha_r} O_p(1).
\end{align*}
 Note that the last two terms are dominated by the first two. Combining results we have
\[ \Bi \tF '\e \e_t/T=O_p(N^{(\alpha_1-2\alpha_r)/2}) + O_p(\frac N T) N^{^{(\alpha_1-2\alpha_r)/2}}+\frac {N^{3/2}} {TN^{\alpha_r}}O_p(1) + N^{\frac 1 2 -\alpha_r} O_p(1). \]
The first term is dominated by the last term since $\alpha_1 \le 1$, and the second term is dominated by the third term, proving (\ref{eq:Feet}).
$\Box$

\paragraph{Proof of Lemma \ref{lem:H3=H0}:}
From
\begin{eqnarray*}
\tF \B-\Fo \B \bar \bH_{NT} & = &\frac{\bm e\bm \Lambda^0  \bm  F^{0^\prime}\tF \Bi}{T}(N \B^{-2} \DNTr^2)^{-1}
+\frac{\bm e \bm e^{\prime}\tF \Bi}{ T} (N \B^{-2} \DNTr^2)^{-1}
= a+b
 \end{eqnarray*}
 Right multiply $\Bi$,
 we have
$\tF =\bm F^0 \bH_{NT}+(a+b)\B^{-1}$ and thus
$\tF'\tF =\tF'\bm F^0 \bH_{NT}+\tF'(a+b)\B^{-1}$.
Dividing by $T$ and using $\tF'\tF/T=I_r$, we obtain
\begin{equation*}
I_r- \bm H_{NT,3}^{-1} \bm H_{NT}= \frac 1 T \tF'(a+b) \Bi =\frac 1 T \tF'a \Bi +\frac 1 T \tF' b \Bi=c +d,
\end{equation*}
where $ \bm H_{NT,3}= (\tF'\Fo/T)^{-1}$.
  Substituting in the expression for $a$, we have
\begin{align*} c  =  \frac 1 T \tF' \frac{\bm e\bm \Lambda^0  \bm  F^{0^\prime}\tF \B^{-2}}{T}(N \B^{-2} \DNTr^2)^{-1}
 & \le  \|\frac {\tF} {\sqrt{T}} \| \cdot \|\frac{\bm e\bm \Lambda^0 \Bi}{\sqrt{T}} \| \cdot \|\B \bm  F^{0^\prime}\tF \B^{-1}\| \cdot \|\Bi\|  O_p(1) \\
&= \|\Bi\| O_p(1)= O_p(N^{-\alpha_r/2} ).
\end{align*}
Substituting the expression for $b$, we have
\begin{align*} d = &   \frac{\tF' \bm e \bm e^{\prime}\tF \B^{-2}}{ T^2} (N \B^{-2} \DNTr^2)^{-1}
\le   \|\bm e\|^2_{sp} \|\tF\|^2 \cdot \|\B^{-2}\| T^{-2} O_p(1) \le \max\{N,T\} N^{-\alpha_r}T^{-1} O_p(1).
\end{align*}
Thus,
\begin{equation} \label{eq2:H3=H0} I_r-  \bm H_{NT,3}^{-1} \bm H_{NT} = O_p(N^{-\alpha_r/2})+ \max\{N,T\} N^{-\alpha_r}T^{-1} O_p(1) \end{equation}
Now  multiply $\bm H_{NT,3}$ on each side of above
\begin{align*} \bm H_{NT,3}-  \bm H_{NT} &  =\bm H_{NT,3} \Big[ O_p(N^{-\alpha_r/2})+ \max\{ N,T\} N^{-\alpha_r}T^{-1} O_p(1)\Big]
\end{align*}
Note that  $ \|\bm H_{NT,3}\| \le N^{(\alpha_1-\alpha_r)/2} O_p(1) $ because we can rewrite
$\bm H_{NT,3} = (\Bi \B \tF'\Fo/T \Bi \B)^{-1}= \Bi ( \B \tF'\Fo \Bi/T)^{-1} \B $, and hence
$\|\bm H_{NT,3}\| \le \|\Bi\|  \cdot  O_p(1) \cdot\|\B\| = N^{(\alpha_1-\alpha_r)/2} O_p(1) $.
It follows that
\[ \bm H_{NT,3}-  \bm H_{NT}  = N^{\frac 1 2 \alpha_1 -\alpha_r}O_p(1)+\frac { N^{1+\frac 1 2 (\alpha_1-3\alpha_r)} }  {T} O_p(1)+  N^{\frac 1 2 (\alpha_1-3\alpha_r)}  O_p(1)  \]
which proves part (i) of the lemma. To prove part (ii), left multiply the above equation by $\bm H_{NT,3}^{-1}$, we have
\[ I_r -  \bm H_{NT} \bm H_{NT,3}^{-1}  =\Big[ N^{\frac 1 2 \alpha_1 -\alpha_r}O_p(1)+\frac { N^{1+\frac 1 2 (\alpha_1-3\alpha_r)} }  {T} O_p(1)+  N^{\frac 1 2 (\alpha_1-3\alpha_r)} O_p(1)\Big] \bm H_{NT,3}^{-1}  \]
But note that $\bm H_{NT,3}^{-1}= \tF'\Fo/T$ so $ \| \bm H_{NT,3}^{-1}\| \le \frac{\|\tF\|}{\sqrt{T}} \frac{ \| \Fo\| }{\sqrt{T}} =O_p(1)$, proving part (ii). $\Box$

\newpage
\begin{table}
\caption{DGP 1}
\label{tbl:table1}
\begin{center}

\begin{tabular}{ll|cccccccccc}  \\ \hline
$ N$ & $T$ & $R^2(\widetilde F_1)$ &  $R^2(\widetilde \Lambda_1)$ & $R^2(\widetilde F_2)$ &  $R^2(\widetilde \Lambda_2)$ &  $R^2(\widetilde F_3)$    & $R^2(\widetilde \Lambda_3) $ &  $M(\tF)$ & $M( \tL)$ & $\bar \rho(\bm C,\tilde{\bm C})$ \\ \hline

\input mc1.out
\hline
\
\end{tabular}
\end{center}Notes: $R^2(\tilde F_j)$ is the $R^2$ from a regression of the PC estimate $\tilde F_j$ on $F^0_1, F^0_2, F^0_3$. $M(\tilde F)$ is the multivariate correlation between $\tilde F$ and $F^0$, and $\bar\rho(\tilde C,C)=\frac{1}{N}\sum_{i=1}^N \rho_i(\tilde C, C^0)$, where $\rho_i$ is the correlation between $\tilde C_i$ and $C_i^0$.
\end{table}
\newpage
\begin{table}
\caption{DGP 2}
\label{tbl:table2}

\begin{center}

\begin{tabular}{ll|cccccccccc} \\ \hline
$ N$ & $T$ & $R^2(\tilde F_1)$ &  $R^2(\tilde \Lambda_1)$ & $R^2(\tilde F_2)$ &  $R^2(\tilde \Lambda_2)$ &  $R^2(\tilde F_3)$    & $R^2(\tilde \Lambda_3) $ &  $M(\tF)$ & $M( \tL)$ & $\bar \rho(\bm C_i,\tilde{\bm C_i})$ \\ \hline

\input mc2.out
\end{tabular}
\end{center} See Table 1 footnotes.
\end{table}

\begin{table}
\caption{DGP2 with  $\Sigma_F \ne I_r$}
\label{tbl:table3}

\begin{center}

\begin{tabular}{ll|cccccccccc} \\ \hline
$ N$ & $T$ & $R^2(\tilde F_1)$ &  $R^2(\tilde \Lambda_1)$ & $R^2(\tilde F_2)$ &  $R^2(\tilde \Lambda_2)$ &  $R^2(\tilde F_3)$    & $R^2(\tilde \Lambda_3) $ &  $M(\tF)$ & $M( \tL)$ & $\bar \rho(\bm C_i,\tilde{\bm C_i})$ \\ \hline

\input mc3.out
\end{tabular}
\end{center}  See Table 1 footnotes.
\end{table}

\newpage
\begin{figure}[ht]
\caption{DGP 1}
\label{fig:fig1}
\includegraphics[width=7.0in,height=3.50in]{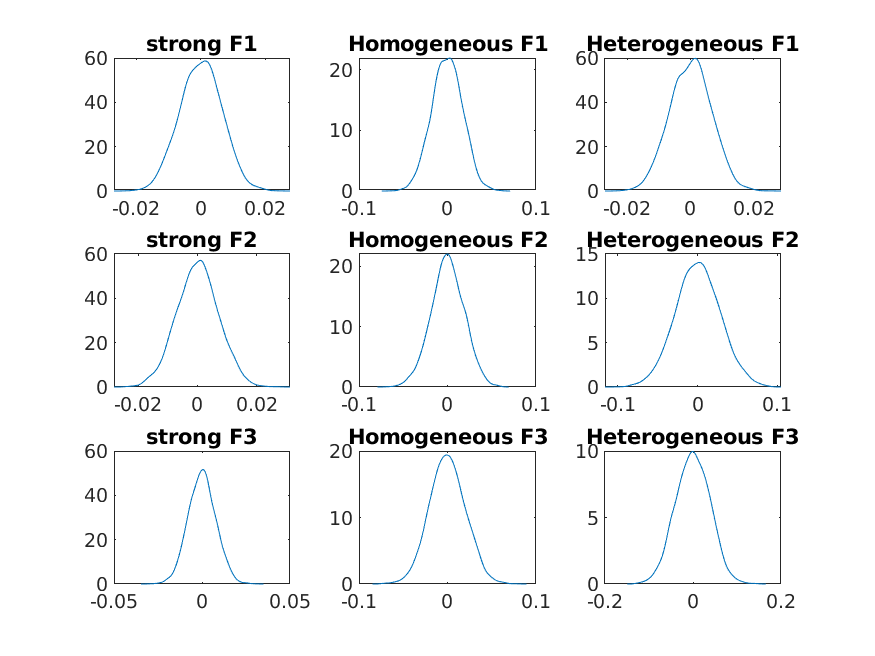}
\includegraphics[width=7.0in,height=3.50in]{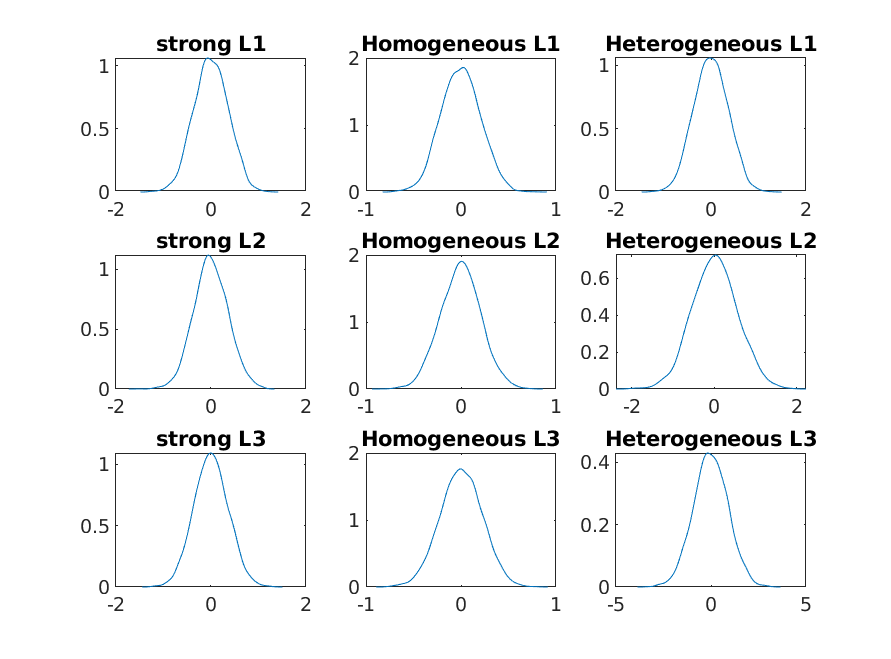}
{\footnotesize Notes: The `Fj 'plots display residuals from regressing $\hat F_{jt}$ on $F^0$ for $t=100$. The `Lj' plots display residuals from  regression $\hat \Lambda_i$ on $\Lambda^0$ for $i=50$. $F^0$ and $\Lambda^0$ are generated using DGP1.}
\end{figure}

\newpage
\begin{figure}[ht]
\caption{DGP 2}
\label{fig:fig2}
\includegraphics[width=7.0in,height=3.50in]{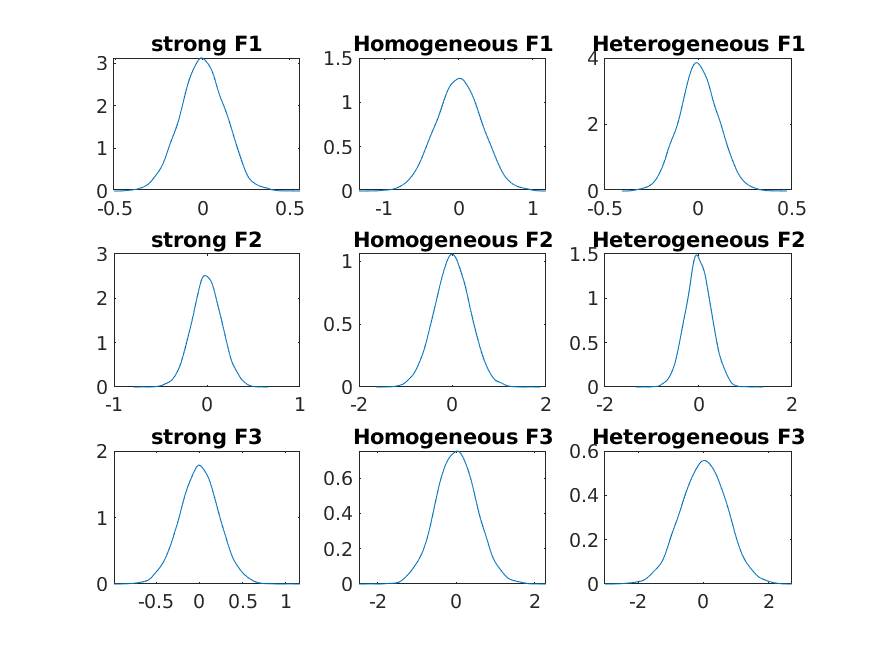}
\includegraphics[width=7.0in,height=3.50in]{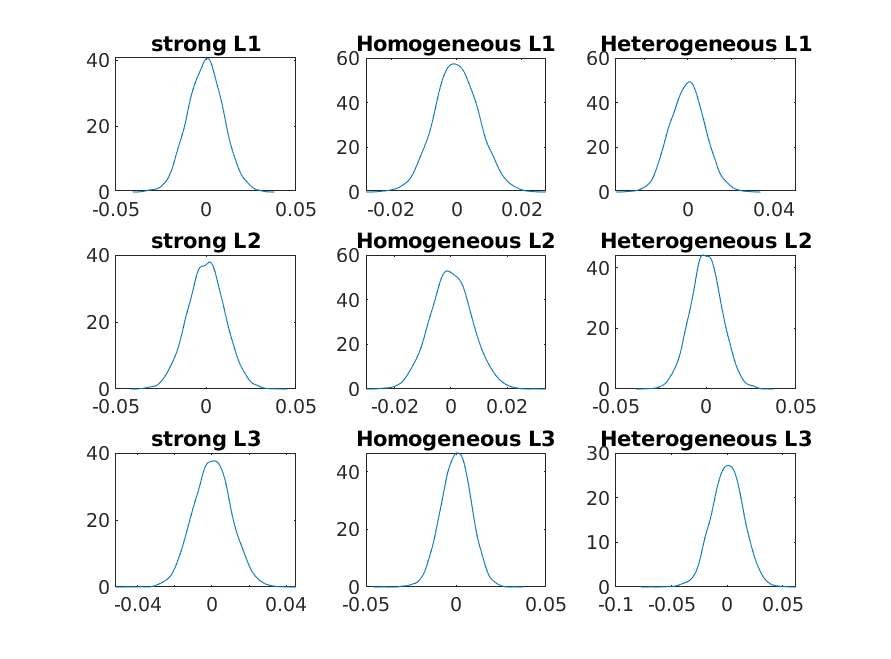}
{\footnotesize
Notes: The `Fj 'plots display residuals from regressing $\hat F_{jt}$ on $F^0$ for $t=100$. The `Lj' plots display residuals from regression $\hat \Lambda_i$ on $\Lambda^0$ for $i=50$. $F^0$ and $\Lambda^0$ are generated using DGP2.}
\end{figure}

\clearpage
\bibliography{factor,metrics,metrics2,bigdata,macro}
\end{document}